\documentstyle[aps,prb,multicol,amssymb,epsf]{revtex}

\begin{document}
\draft
%\tighten
\title{Higher-Order Results for the Relation between
Channel Conductance and the Coulomb Blockade for
Two Tunnel-Coupled Quantum Dots}
\author{John M. Golden and Bertrand I. Halperin}
\address{Department of Physics, Harvard University, Cambridge, MA
  02138}

\date{Submitted 15 March 1996}

\maketitle

\begin{abstract}
\noindent \ \
We extend earlier results on the relation between the dimensionless
tunneling channel conductance $g$ and the fractional
Coulomb blockade peak splitting $f$ for two
electrostatically equivalent dots connected by an arbitrary
number $N_{\text{ch}}$ of tunneling channels with bandwidths $W$
much larger than the two-dot differential charging energy $U_{2}$.
By calculating  $f$ through second
order in $g$ in the limit of weak coupling ($g \rightarrow 0$),
we illuminate the difference in behavior of
the large-$N_{\text{ch}}$ and small-$N_{\text{ch}}$ regimes and
make more plausible extrapolation to the strong-coupling
($g \rightarrow 1$) limit.  For the special case of $N_{\text{ch}}=2$
and strong coupling, we eliminate an apparent ultraviolet
divergence and obtain the next leading term of an expansion
in $(1-g)$.  We show that the results we calculate are
independent of such band structure details as the fraction of
occupied fermionic single-particle states in the weak-coupling
theory and the nature of the cut-off in the bosonized strong-coupling
theory.  The results agree with calculations for metallic
junctions in the $N_{\text{ch}} \rightarrow \infty$ limit and
improve the previous good agreement with recent two-channel
experiments.
\end{abstract}


\begin{multicols}{2}

\narrowtext

\section{Introduction}

The opening of tunneling channels between two
quantum dots leads to an erosion of the individual dots'
Coulomb blockade.~\cite{Review}  For a pair of
electrostatically identical quantum dots (see Fig. 1 for
a schematic view of the double-dot structure), the progress
of this erosion can be chronicled by tracking the splitting of
the Coulomb blockade conductance peaks as
they evolve from doubly degenerate single-dot
conductance resonances to that of nondegenerate double-dot
peaks with twice the original
periodicity.~\cite{Waugh1,Waugh2,Golden1,Matveev3,Matveev4}
For a system in which the tunneling channels can be treated
as having the
same individual conductances and in which
the Coulomb charging energies
are large compared to the single-particle
level spacings but small compared to the tunneling channel
bandwidths, the fractional peak splitting $f$ can be expressed
as a function of two parameters: $N_{\text{ch}}$, the number of
tunneling channels between the two dots, and $g$, the
dimensionless conductance per tunneling channel.
(In this paper, the {\em conductances} indicated are always
{\em dimensionless conductances}, by which we mean
the actual conductance divided by the conductance quantum,
$e^{2}/h$.)

In particular, for weakly coupled dots ($g \rightarrow 0$),
the fractional peak splitting can be expressed perturbatively
as a sum of terms of the
form $a_{m,n} (N_{\text{ch}})^{m} g^{n}$,
where  $1 \leq m \leq n$ and
$a_{m,n}$ is independent of $N_{\text{ch}}$ and $g$.
Previous work~\cite{Golden1,Matveev3,Matveev4} has produced
the leading term in this expansion.  However, as this term is
simply linear in the total interdot tunneling conductance,
$g_{\text{tot}} = N_{\text{ch}} g$, it does not effectively
distinguish between behavior in the
large-$N_{\text{ch}}$ and small-$N_{\text{ch}}$
limits.  To make such a distinction, one must calculate to
second order in $g$, in which case one obtains two sets of
terms, one set proportional to
$N_{\text{ch}} g^2 = g_{\text{tot}}^2/N_{\text{ch}}$
and the other proportional to
$(N_{\text{ch}})^2 g^2 = g_{\text{tot}}^2$.  The
$(N_{\text{ch}})^2$ terms should agree with results from the theory
of metallic junctions, in which the leading terms in a
large-$N_{\text{ch}}$ expansion are
calculated.~\cite{Golubev,Grabert}

In the strong-coupling limit ($g \rightarrow 1$), a
dramatic dependence of the peak splitting on $N_{\text{ch}}$
has already been found.  In the cases of $N_{\text{ch}}=1$
and $N_{\text{ch}}=2$, the leading nontrivial
terms have been found to be proportional to $\sqrt{1-g}$ and
$(1-g) \ln(1-g)$,
respectively,~\cite{Golden1,Matveev3,Matveev4,Matveev2}
and it has been hypothesized~\cite{Molen,Flensberg}
that, for $N_{\text{ch}} > 2$ but finite,
the leading nontrivial term is proportional to
$(1-g)^{N_{\text{ch}}/2}$.  This last
suggestion appears to
correspond to calculations in the large-$N_{\text{ch}}$
limit,~\cite{Zimanyi,Panyukov} where
the effective charging energy has been found to be proportional to
$e^{-g_{\text{tot}}/2}$, which is equivalent to
$(1-g_{\text{tot}}/N_{\text{ch}})^{N_{\text{ch}}/2}$ in the limit
$N_{\text{ch}} \rightarrow \infty$.

Despite the recent progress in study of the strong-coupling limit,
for the case of most direct experimental
interest,~\cite{Waugh1,Waugh2,Molen} $N_{\text{ch}}=2$, the leading-term
calculation fails to be completely satisfactory
for at least three reasons.
The first is that this calculation does not answer the
question of whether the coefficient of \mbox{ $(1-g) \ln (1-g)$ }
is affected by the manner in which the ultraviolet cut-off is imposed
in the low-energy Luttinger liquid theory.~\cite{Golden1}
The second is that the coefficient of the sub-leading term
linear in $(1-g)$ is both unknown and naively infinite.~\cite{Matveev2}
Finally, there is the worry---which also applies to the
weak-coupling result---that, for $N_{\text{ch}}=2$, interpolation
between the solutions for weak- and strong-coupling is
difficult because the respective $f$-versus-$g$
curves do not come especially close.~\cite{Golden1}

This paper addresses these three concerns for the two-channel
problem and also extends earlier results for the general
$N_{\text{ch}}$-channel problem in the limit of
weak coupling.  In so doing, it illuminates the difference
between large-$N_{\text{ch}}$ and small-$N_{\text{ch}}$
behavior for $g \approx 0$, creates a theory that can be
more realistically compared to experimental results for
$N_{\text{ch}} = 2$, and argues for the universality of
the results, which should be independent of the nature
and magnitude of the ultraviolet cut-offs.
Section II presents the $g^2$ extension
of the weak-coupling theory and
checks the result against calculations in the
$N_{\text{ch}} \rightarrow \infty$ limit.  Section III gives the
$(1-g)$ correction to the leading dependence in the
strong-coupling limit for $N_{\text{ch}}=2$ and shows a
plot of the experimental results and revised theoretical
predictions for two-channel interdot junctions.
Section IV argues that the strong coupling results of
Section III are independent of the nature of the way
the ultraviolet cut-off is imposed and do not change when
one allows the fermionic theory to stray from half-filling.
Section V summarizes the results, and Appendices A and
B present technical details of calculations in
Sections II and III, respectively.

\begin{minipage}{3.27truein}
  \begin{figure}[H]
    \begin{center}
      \leavevmode
      \epsfxsize=3.27truein
      \epsfbox{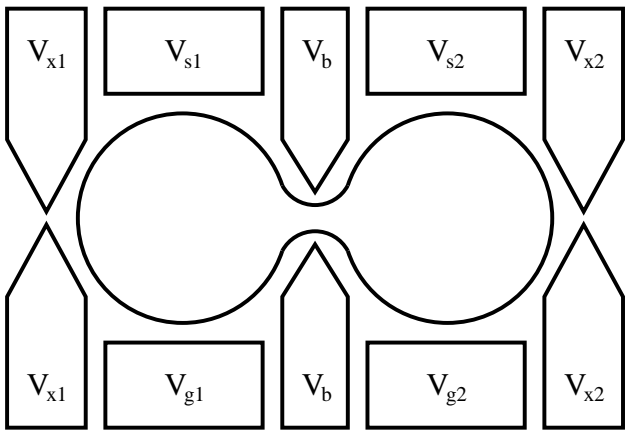}
    \end{center}
    \caption{Schematic diagram for the double dot.  Negative
  potentials are applied to each of the gates to form the
  double-dot structure.  The gate potentials $V_{\text{g1}}$
  and $V_{\text{g2}}$
  control the average numbers of electrons on the dots.  These
  are the potentials that are varied to see the Coulomb blockade.
  $V_{\text{b}}$ controls the rate of tunneling between the dots.
  $V_{\text{x1}}$ and $V_{\text{x2}}$ control the rate of tunneling to
  the adjacent bulk two-dimensional electron-gas (2DEG) leads.
  For calculations of the double-dot energy shifts, tunneling to
  the leads is assumed negligible compared to tunneling between
  the two dots.  In measuring the total channel conductance $G_{\text{tot}}$,
  however, the potentials $V_{\text{xi}}$ are
  turned off so that each dot is strongly connected to its lead.
  The side-wall potentials $V_{\text{s1}}$ and $V_{\text{s2}}$ are
  fixed.}
    \label{fig:dots}
  \end{figure}
  \smallskip
\end{minipage}

\section{The Weak-Coupling Limit for Arbitrary
           $N_{\text{ch}}$ }

For weakly coupled quantum dots, we use a model
``site-to-site'' hopping Hamiltonian~\cite{Golden1} and
calculate perturbatively in the tunneling term
$H_{T}$:
\begin{eqnarray}
H_{\ }  & = & H_{K} + H_{C} \, ,  \nonumber \\
H_{K} & = & \sum_{i=1}^{2} \sum_{\sigma} \sum_{\bf{k}}
          \epsilon_{\bf{k}} \hat{n}_{i \bf{k} \sigma} \, ,
      \nonumber  \\
H_{C} & = & U_{2} (\hat{n} - \rho/2)^{2} \, ,
                        \nonumber  \\
H_{T}  & = & \sum_{\sigma} \sum_{\bf{k_1} \bf{k_2}}
       t(c^{\dagger}_{2 \bf{k_2} \sigma}
     c_{1 \bf{k_1} \sigma} + \mbox{H.c.}) \, .
\label{eq:weakham}
\end{eqnarray}
As in Ref. 4, in these equations,
$i$ is the dot index; $\sigma$ is the channel index;
$\bf{k}$ is the index for all internal degrees of freedom
not included in the channel index; $H_{C}$ is the
part of the electrostatic potential energy that is
affected by interdot tunneling; $\hat{n}$ is half
the difference in dot occupation numbers,
$\hat{n}=(\hat{n}_{2}-\hat{n}_{1})/2$;
$\rho$ is a differential gate voltage parameter and is
restricted to values between 0 and 1 (as permitted
by the system's unit periodicity); and $U_{2}$ is the
differential charging energy, which, for
electrostatically equivalent dots, is given by the formula
$U_{2}=e^{2}/(C_{\Sigma}+2 C_{\text{int}})$, where
$C_{\text{int}}$ is the interdot capacitance and
$C_{\Sigma}$ is the total single-dot capacitance minus
the interdot capacitance.  If the dots are not
electrostatically equivalent, the formula for $U_{2}$
and the definition of $\rho$ are more
complicated.~\cite{Matveev3,Matveev4}  However,
the model is still applicable, and the results for
$f_{\rho}$ can still be used to obtain the
peak splitting.

These calculations are made palatable by assuming that
$U_{2}$ is  much smaller than the
tunneling-channel bandwidth $W$ yet much greater than
the average intrachannel level-spacing $\delta$:
\begin{math}
W \gg U_{2} \gg \delta.
\end{math}
This assumption leaves us with a theory that we can
consider to be in the continuum limit and that we can
hope to be independent of ultraviolet cut-offs.
As the bandwidth is presumably of the order
of the Fermi energy $\epsilon_{F}$, these assumptions
are reasonable for the micrometer-sized dots of
Waugh {\em et al.},~\cite{Waugh1,Waugh2} for which
$\epsilon_{F} \approx 10$ meV,
$U_{2} \approx 400 \mbox{ } \mu$eV, and
$\delta \approx 30 \mbox{ } \mu$eV.

As in Ref. 4, our primary goal is to
calculate the fractional peak splitting $f$---i.e.,
the ratio of the distance between split
Coulomb blockade subpeaks for a given $g$ and their maximal
separation in the strong-coupling ($g \rightarrow 1$) limit.
It was shown in Ref. 4 that,
if the total number of electrons on the two dots
is assumed even, the
problem of solving for $f$ is a corollary to
the problem of solving for a more general
quantity $f_{\rho}$, which characterizes the ground-state
energy of the double-dot when the difference between the
external potentials applied to the dots is nontrivial
and the total number of electrons on the two dots is
fixed and even.
Recall the equation for $f_{\rho}$:
\begin{equation}
f_{\rho} = \frac{\Delta_{0} - \Delta_{\rho}}{U_{2} \rho^2/4} \, ,
\label{eq:frho}
\end{equation}
where $\Delta_{\rho}$ is the shift in the ground
state energy induced by tunneling at a given
value of the gate voltage parameter $\rho$
and $U_{2} \rho^2/4$ is the
difference between the unperturbed ground state
energies for the given $\rho$ and $\rho = 0$.
In Ref. 4, it was shown that, for
symmetric dots,
\begin{equation}
f = f_{\rho=1} \, .
\end{equation}
In the same work, it was determined that $f_{\rho}$
exhibits the following
leading behavior as $g \rightarrow 0$:
\begin{eqnarray}
f^{(1)}_{\rho} & = & \frac{N_{\text{ch}}g}{\pi^{2} \rho^2}
  [(1-\rho) \ln (1-\rho) + (1+\rho) \ln (1+\rho) \nonumber \\
               &   &  \mbox{\hspace{0.6in}} + {\cal O}(\rho^{2} / \psi)] \, ,
\label{eq:weak1rho}
\end{eqnarray}
where $\psi = W/U_{2} \gg 1$.  Thus, the corresponding
leading behavior for $f$ is
\begin{equation}
f^{(1)} = \frac{2 \ln 2}{\pi^2} N_{\text{ch}} g
        + {\cal O}(g/\psi,g^2) \, .
\label{eq:weak1}
\end{equation}

Extending perturbation theory beyond this result---i.e.,
beyond first-order in $g$---
requires some laborious computation.
The next-leading contributions come from two sources.
The first, which we shall call $f^{(2A)}_{\rho}$, arises
from a combination of
the second-order energy shift that has already been
calculated and the second term in
the formula that relates the tunneling amplitude
$t$ to the channel conductance $g$.  [The first term in
this formula was used to derive Eq.~(\ref{eq:weak1rho}).]
The second source of $g^2$ terms, $f^{(2B)}_{\rho}$, is
the shift in the ground-state energy
provided by terms that are fourth-order in $t$.

The first contribution is relatively easy to calculate.
The equation for $g$ in terms of $t$ has been derived for
half-filling in Ref. 14 and for arbitrary filling
in Ref. 4.  In the latter calculation, the
system is assumed to have a constant density of states
between single-particle energies $\epsilon_0$ and
$(\epsilon_0 + W)$, the density of states being zero
elsewhere.  The system's level of ``filling'' is then
characterized by the filling fraction
$F = (\epsilon_F - \epsilon_0)/W$, where $\epsilon_F$ is
the Fermi energy.  In accordance with the half-filling
result, one then finds the following:
\begin{equation}
g  = \frac{4 \chi}{|1 + (1+ {\em i} \eta)^{2}  \chi|^{2}} \, \, ,
\label{eq:g-of-t}
\end{equation}
where $\chi = (\pi t / \delta)^{2}$ and
$\eta = (1/\pi)\ln[F/(1-F)]$.
Inverting this expression, one discovers that
\begin{equation}
\frac{t^{2}}{\delta^{2}} = \frac{g}{4 \pi^2}
     \left[1 + \frac{1-\eta^{2}}{2} g + {\cal O}(g^{2}) \right] \, .
\label{eq:t-of-g}
\end{equation}
Consequently, our first $g^{2}$ term
is equal to the right side of
Equation~(\ref{eq:weak1rho}) multiplied by \mbox{$(1-\eta^{2})g/2$}:
\begin{eqnarray}
f^{(2A)}_{\rho} & = & (1-\eta^{2})
    \frac{N_{\text{ch}}g^{2}}{2 \pi^{2} \rho^2}
    [(1-\rho) \ln (1-\rho)  \nonumber \\
                &   & \mbox{\hspace{0.4in}} + (1+\rho) \ln (1+\rho)
       + {\cal O}(\rho^{2} / \psi)] \, .
\label{eq:weak2A}
\end{eqnarray}
This term is of the expected form
$a_{\text{1,2}}^{(2A)} N_{\text{ch}} g^{2}$,
where $a_{\text{1,2}}^{(2A)}$ is a function of $\rho$.

On the other hand, $a_{\text{1,2}}^{(2A)}$ is dependent
on the filling fraction $F$, a fact which appears
to imperil our dreams of a theory that is
universal in that it is insensitive to
the details of the high-energy behavior
(including whether, for example, certain high-energy
states exist and therefore have a role in
determining the filling fraction $F$).  We shall see,
however, that the $F$-dependence of $f^{(2A)}_{\rho}$
actually serves our end, for it exactly
cancels the $F$-dependence of $f_{\rho}^{(2B)}$.
As a result, we can further conclude that,
through second order in the channel conductance $g$,
expression of the fractional peak splitting in terms
of the channel conductance is not only convenient for
comparison with experiment but is also
necessary and sufficient for constructing a result that can
be hoped to be universal.

To support this claim, we must actually determine
the value of $f^{(2B)}_{\rho}$.  Sadly, it cannot be obtained
as effortlessly as $f^{(2A)}_{\rho}$.
There are 24 separate terms that contribute to
the fourth-order energy shift.
One 12-member subset consists of
terms proportional to $(N_{\text{ch}})^2$; the second
consists of those simply linear in $N_{\text{ch}}$.
All but four of the twenty-four terms correspond to a
specific series of four tunneling events that begin
and end with the double-dot system's unperturbed ground state.
The remaining four, which belong to the $(N_{\text{ch}})^2$ subset,
correspond to the fourth-order terms in
Rayleigh-Schr\"{o}dinger perturbation theory that are
products of the second-order energy shift and a
propagator squared.  These four have been described by Grabert
as {\em diagrams with insertions}.~\cite{Grabert}

In general, the nature of the twenty-four fourth-order terms
is most digestibly summarized via a diagrammatic representation
that looks essentially like one of time-ordered single-particle
diagrams (see Fig. 2).  Despite the superficial single-particle
nature of this representation, it is important to remember
that the propagators that enter into the energy calculations
are the propagators for the entire double-dot system, which
depend upon both the tunneling particles' individual
kinetic energies and the system's multiparticle potential
energy.  The presence of the multiparticle potential energy
makes it impossible to reduce the calculation to the
normal Feynman diagrams,
for which one can write the problem entirely in terms of
single-particle propagators.  The presence of exchange terms,
which do not appear among the diagrams proportional to
$(N_{\text{ch}})^2$, makes a pseudo-single-particle representation
necessary.

Within this time-ordered perturbation theory scheme,
the individual fourth-order terms are plagued by both ultraviolet
and infrared divergences.  Every term is divergent
as the bandwidth goes to infinity and four of the $(N_{\text{ch}})^2$
terms are divergent as $\rho \rightarrow 1$.  (A different set
of four is divergent as $\rho \rightarrow -1$.)
{}From the result for $f^{(1)}_{\rho}$,
we might hope to cancel
the ultraviolet divergences and to obtain an answer for the
ground-state energy that is infrared-singular but not
infrared-divergent.  Indeed, as Grabert has
noted,~\cite{Grabert} the ultraviolet
divergences of the $(N_{\text{ch}})^2$ terms must drop out
since, in the limit $U_{2} \rightarrow 0$, these terms
correspond to disconnected diagrams or insertion diagrams
that exactly cancel one another and
thus do not appear as Feynman diagrams.
In contrast, the $N_{\text{ch}}$ diagrams do have nontrivial
Feynman-diagram analogs.  As a whole, they correspond
to a single totemic Feynman diagram---an individual ring
marked by four

\begin{minipage}{3.27truein}
  \begin{figure}[H]
    \begin{center}
      \leavevmode
      \epsfxsize=2.5truein
      \epsfbox{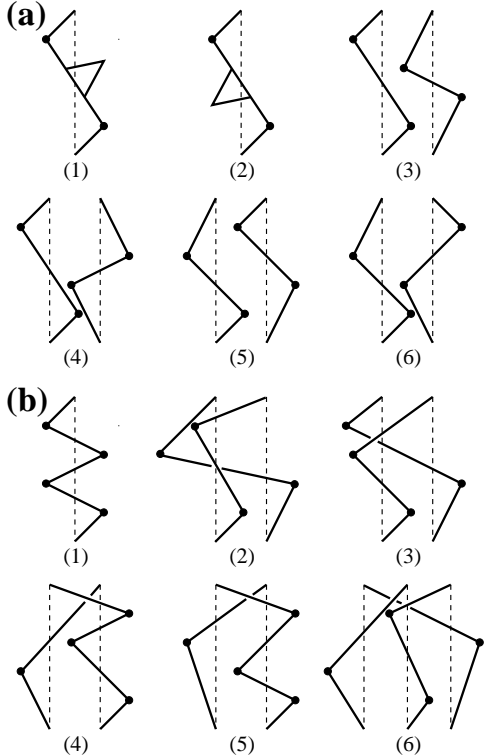}
    \end{center}
    \caption{Diagrams for half of the (a) fourth-order, $(N_{\text{ch}})^2$
  terms and (b) fourth-order, $N_{\text{ch}}$ terms.  The remaining terms
  are represented by diagrams that are mirror images of these.
  A vertical dashed line is drawn for each of the $m$ particles
  that tunnels at least once from one dot to the other.  This
  line stands for the corresponding particle's initial state,
  a state that must be filled at the end of the four tunneling
  events in order to recover the unperturbed ground state from
  which the system starts.  A particle begins at the bottom of
  its vertical initial-state line.  Particles in dot 1 propagate
  upward and rightward.  Particles in dot 2 propagate
  upward and leftward.  A tunneling event for a particle is
  signaled by a solid dot that coincides with a bend in the
  particle-propagation path.  Each particle must end on one of the
  dashed vertical lines, meaning that it ends in the
  single-particle state that corresponds to that line.
  {\em Insertions} (see Sec. II) are
  represented by triangles that project off a single-particle
  propagation line.  If the projection points up, the insertion
  corresponds to the term in the second-order energy shift for which
  a particle tunnels off the dot occupied by the propagating
  particle.  If the projection points down, the insertion
  corresponds to the second-order term for which a particle
  tunnels onto the dot occupied by the propagating particle.
  In the absence of exchange, all particles
  end on their own {\em initial-state lines}.  A two-particle
  exchange carries a minus sign and results in each of two
  particles ending on the other's initial-state line.
  Three-particle exchange carries no sign (alternatively,
  one can view it as carrying two canceling minus signs)
  and results in each of three particles ending on one of the
  others' initial-state lines.}
    \label{fig:raysch}
  \end{figure}
  \smallskip
\end{minipage}

\noindent  tunneling events.  The ultraviolet divergences
of these diagrams are therefore expected to be persistent
but irrelevant because we are interested only
in the relative shift between the ground-state energies
for arbitrary $\rho$ and for $\rho = 0$
(recall Eq.~\ref{eq:frho}).
Accordingly, we expect that, when one subtracts the fourth-order
shift for $\rho=0$ from that for arbitrary $\rho$, the
fourth-order terms produce a result that is neither
ultraviolet- nor infrared-divergent
but is infrared-singular as $|\rho| \rightarrow 1$.
A brief summary of the actual calculation of these terms follows.
Those interested in more detail are invited to peruse
Appendix~A, which offers a fuller description of the
calculation of the $(N_{\text{ch}})^2$ diagrams and a step-by-step
computation of the contribution from one representative $N_{\text{ch}}$
term.

For the less scrupulous, there are still a few facts
worthy of note.  A prominent feature of the
fourth-order calculation is that
each term involves the integration over four energy variables
($\epsilon_i$, where $i$ ranges from 1 to 4)
of a product of three propagators.
In the $(N_{\text{ch}})^2$ diagrams, the energy variables
``pair off'':  $\epsilon_1$ and $\epsilon_3$ only appear
as parts of the combination
$\epsilon_{\text{I}} = (\epsilon_3 - \epsilon_1)$, and
$\epsilon_2$ and $\epsilon_4$ only appear as parts of the
combination $\epsilon_{\text{II}} = (\epsilon_4 - \epsilon_2)$.
As a result, calculation of these terms reduces to the
performance of double integrations
over $\epsilon_{\text{I}}$ and $\epsilon_{\text{II}}$---
albeit with a nontrivial density of states.

The $N_{\text{ch}}$ diagrams cannot be handled in this
way, for they involve particle exchanges that
frustrate any desire to pair off the energy variables.
The quadruple integration over the $\epsilon_i$ cannot
be eluded.  It can, however, be expedited by differentiating
twice with respect to $\rho$ while integrating out the energy
variables and, then, integrating twice with respect
to $\rho$ in the end.  One might worry about the fact that,
by differentiating twice with respect to $\rho$, one has
lost knowledge of terms constant and linear in $\rho$.
However, these terms are unimportant.
As noted in Ref.~4,
the ground-state energy (perturbed or unperturbed) is symmetric
in $\rho$.  Therefore, terms linear in
$\rho$ must cancel out of the fourth-order energy shift when
all the terms are summed.  Constant terms are similarly
negligible since, as usual, we are only interested in the
relative energy shift $\Delta_{\rho} - \Delta_{0}$.

After the aforementioned tricks for calculating the
$(N_{\text{ch}})^2$ and $N_{\text{ch}}$ diagrams have
been used, the only real wrinkles that remain are
integrals of the form
\begin{displaymath}
{\cal P} \int_{0}^{R \psi} \! dx \, \frac{\ln (x + B)}{(x + A)} \, ,
\end{displaymath}
where $0 \leq |A| < B$, $R$ is either $F$ or $(1-F)$,
and, as before, $\psi = W/U_{2}$.
The symbol ${\cal P}$ indicates that, for $A < 0$, only
the principal value of the integral is calculated.
These integrals can be done by rewriting the argument
of $\ln (x+B)$ as $[(x+A)+(B-A)]$ and Taylor-expanding
about $(x+A)$ for $(B-A) < |x+A|$ and about $(B-A)$
for $(B-A) > |x+A|$.  The result of such an integration
may be sensitive to whether the system is below
half-filling [$F<(1-F)$], at half-filling [$F=(1-F)$],
or above half-filling [$F>(1-F)$].  However, the system
as a whole has particle-hole symmetry, so
one expects that the final result---once all the terms
are summed---is symmetric under exchange of $F$ and
$(1-F)$.  If there is no jump
discontinuity when the system is precisely half full,
the result for $F<(1-F)$ should determine the answer for
all ``finite'' $F$, by which we mean
all $F$ such that $F \psi, (1-F) \psi \gg 1$.
This thesis has been explicitly confirmed.

Indeed, the $(N_{\text{ch}})^2$ part of the fourth-order
relative energy shift
is found to be independent of the filling fraction.
Its contribution to $f_{\rho}$ has a rather lengthy
explicit form:
\begin{eqnarray}
f^{(2B)}_{\rho,(N_{\text{ch}})^2} & = &
\frac{(N_{\text{ch}})^2 g^2}{4 \pi^4 \rho^2}
 \{ - \frac{\pi^2}{6} \rho^2  +4(1-\rho)\ln (1-\rho) \nonumber \\
 & & \mbox{}  +\frac{1-\rho^2}{2} \ln^2 (1-\rho)
      -2(2-\rho) \ln[2(2-\rho)] \nonumber \\
 & & \mbox{}
              + \frac{1+\rho^2}{2} \ln(1+\rho) \ln(1-\rho) \nonumber \\
 & & \mbox{}
         -2 \ln (3-\rho) \ln(1-\rho) + \ln^2 (3-\rho) \nonumber \\
 & & \mbox{}
 + \frac{(3-\rho)(1-\rho)}{2} [\ln(1-\rho) - \ln(3-\rho)]^2 \nonumber \\
 & & \mbox{} - (5 - 4\rho + \rho^2)
                    \sum_{n=1}^{\infty} \frac{(-1)^{n+1}}{n^2}
                    \left(\frac{1-\rho}{3-\rho} \right)^{n}
 %& & \mbox{} - 2\left[1 + \frac{(3-\rho)(1-\rho)}{2} \right]
 %      \sum_{n=1}^{\infty} \frac{(-1)^{n+1}}{n^2}
 %                   \left(\frac{1-\rho}{3-\rho} \right)^{n}
 \nonumber \\
 & & \mbox{} -\frac{5 \ln^2 3}{2} + 8 \ln 2
       + 5 \kappa  \nonumber \\
 & & \mbox{} +[\rho \rightarrow -\rho] \mbox{\ }  \} \, ,
\label{eq:f2Bsqr}
\end{eqnarray}
where
the contents of the last pair of brackets indicate that one
sums over all the terms in the curly braces again after replacing
$\rho$ with $-\rho$
and the quantity $\kappa$ is given by
\begin{equation}
\kappa = \sum_{n=1}^{\infty} \frac{(-1)^{n+1}}{n^2}
             \left( \frac{1}{3} \right)^{n} \, .
\label{eq:kappa}
\end{equation}

It has been confirmed numerically that the
derivative of $f^{(2B)}_{\rho,(N_{\text{ch}})^2}$
agrees with a calculation by Grabert in the
$N_{\text{ch}} \rightarrow \infty$ limit.~\cite{Grabert}
More precisely, the derivative of this term with respect to $\rho$
corresponds exactly to Grabert's calculation of the leading
corrections to the ground-state expectation value of $\hat{n}$.
The value at $\rho = 0$ of the second derivative of this term with
respect to $\rho$ (and of the first derivative with respect to $\rho$
of Grabert's $\hat{n}$ correction) should agree with the
large-$N_{\text{ch}}$ ``effective capacitance''
of Golubev and Zaikin.~\cite{Golubev}  However,
this is found not to be so, the $g^2$ correction calculated by
Golubev and Zaikin being nearly a factor of 20 smaller than that
computed here.  The present calculation therefore
provides an important check on the large-$N_{\text{ch}}$ calculations,
resolving an apparent contradiction in the literature.

There are no corroborating calculations for the terms that
are linear in $N_{\text{ch}}$ as these are negligible in the
large-$N_{\text{ch}}$ limit.  However, knowing that $f = 1$ when
$g = 1$ and that $f^{(1)} \approx (0.14) N_{\text{ch}} g$,
one might conjecture that the sign of the $g^2$ term changes
from positive to negative when $N_{\text{ch}}$ is of order $10$.
With respect to the expansion of $f$, such a crossover would
imply that the coefficient of $N_{\text{ch}} g^2$ is positive and
approximately 10 times the size of the negative coefficient
of $(N_{\text{ch}})^2 g^2$.

To check this conjecture, we need to know the value of the
fourth-order, linear-in-$N_{\text{ch}}$ contribution to the
fractional peak splitting.  Our results for this quantity are
\begin{eqnarray}
f^{(2B)}_{\rho,N_{\text{ch}}} & = &
\frac{N_{\text{ch}} g^2}{4 \pi^4 \rho^2}
 \{ 2 \pi^2 \eta^2 (1-\rho)\ln(1-\rho) + \frac{4}{3} (1-\rho)\ln^3(1-\rho)
     \nonumber \\
 & & \mbox{} - 2(2 + \ln 2)(1-\rho)\ln^2(1-\rho)  \nonumber \\
 & & \mbox{}+ 4 \left(2 + \ln 2 - \ln 2 \ln 3 -\kappa
                 + \frac{\pi^2}{3} \right)(1-\rho)\ln(1-\rho)   \nonumber \\
 & & \mbox{}-2\left(\frac{\pi^2}{3} - 2\kappa - 2\ln 2\ln 3 \right)
                             [ (3-\rho)\ln (3-\rho) - 3 \ln 3]
                                                \nonumber \\
 & & \mbox{} -\frac{2}{3}[ (3-\rho)\ln^3(3-\rho)
                               -3(3-\rho)\ln^2(3-\rho)
                                                 \nonumber \\
 & & \mbox{\hspace{0.3in}}+6(3-\rho)\ln(3-\rho) - 3\ln^3 3 + 9\ln^2 3
                        - 18\ln 3]                \nonumber \\
 & & \mbox{}-2\ln 2 [ (3-\rho)\ln^2 (3-\rho) - 2(3-\rho)\ln(3-\rho)
                          \nonumber \\
 & & \mbox{\hspace{0.3in}}
                         - 3\ln^2 3 + 6\ln 3 ]      \nonumber \\
 & & \mbox{}-2\sum_{i=1}^{11} {\em A}_{i}(\rho)
                              \nonumber \\
 & & \mbox{} +[\rho \rightarrow -\rho] \mbox{\ }
        \} \, ,
\label{eq:f2Blin}
\end{eqnarray}
where $\kappa$ is given by Eq.~(\ref{eq:kappa}),
$\eta = (1/\pi)\ln [F/(1-F)]$,
and the ${\em A}_{i}(\rho)$ are defined below:
\begin{eqnarray}
{\em A}_{1}(\rho) & = &
    -\int_{0}^{\rho} \! dx \, \frac{(\rho-x)\ln^2(3-x)}{1-x} %\, ,
                               \nonumber \\
{\em A}_{2}(\rho) & = & 2\int_{0}^{\rho} \! dx \,
                     \frac{(\rho-x)\ln(3-x)\ln(1-x)}{1-x} %\, ,
                               \nonumber \\
{\em A}_{3}(\rho) & = & -2\ln 2 \int_{0}^{\rho} \! dx \,
                    \frac{(\rho-x)\ln(3-x)}{1-x} %\, ,
                               \nonumber \\
{\em A}_{4}(\rho) & = & 2(1-\rho)\int_{0}^{\rho} \! dx \,
                    \ln \! \left( \frac{2-x}{3-x} \right)
                    \ln \! \left( \frac{1-\rho}{1-x} \right)
                         \nonumber \\
 & & \mbox{\hspace{0.6in}} \times
                    \left( \frac{1}{3-x} - \frac{1}{1-x} \right) %\, ,
                               \nonumber \\
{\em A}_{5}(\rho) & = & 2\int_{0}^{\rho} \! dx \,
 (\rho-x) \ln \! \left(\frac{2-x}{3-x} \right)
        \left( \frac{1}{3-x} - \frac{1}{1-x} \right) % \, ,
                               \nonumber \\
{\em A}_{6}(\rho) & = & \int_{0}^{\rho} \! dx \,
                         \frac{(\rho-x)\ln^2(1-x)}{3-x} % \, ,
                               \nonumber \\
{\em A}_{7}(\rho) & = & -2\int_{0}^{\rho} \! dx \,
     \frac{(\rho-x)\ln(3-x) \ln(1-x)}{3-x} % \, ,
                               \nonumber \\
{\em A}_{8}(\rho) & = & -2\ln 2\int_{0}^{\rho} \! dx \,
            \frac{(\rho-x)\ln(1-x)}{3-x} % \, ,
                               \nonumber \\
{\em A}_{9}(\rho) & = & -2(3-\rho) \int_{0}^{\rho} \! dx \,
                    \ln \! \left( \frac{2-x}{3-x} \right)
                    \ln \! \left( \frac{3-\rho}{3-x} \right)
                         \nonumber \\
 & & \mbox{\hspace{0.6in}} \times
             \left( \frac{1}{3-x} - \frac{1}{1-x} \right) % \, ,
                               \nonumber \\
{\em A}_{10}(\rho) & = & -2\int_{0}^{\rho} \! dx \,
           (\rho-x) \ln \! \left(\frac{2-x}{3-x} \right)
                               \nonumber \\
 & & \mbox{\hspace{0.6in}} \times
                      \left( \frac{1}{3-x} - \frac{1}{1-x} \right) % \, ,
                               \nonumber \\
{\em A}_{11}(\rho) & = & -\int_{0}^{\rho} \! dx \, \ln^2(1-x) \ln(1+x) % \, .
\label{eq:Ai}
\end{eqnarray}

The characterization of the fourth-order energy shift is now
essentially complete.  The result is more unwieldy than
we would like.  However, there are a few highlights that
are easy to draw out.  As expected, the fourth-order
shift is neither ultraviolet- nor infrared-divergent but
is singular as $|\rho| \rightarrow 1$, the leading singularities
being in agreement with an earlier calculation by Glazman
and Matveev.~\cite{Matveev1}  In addition and
quite gratifyingly, the solution is independent of the
filling fraction $F$.  As discussed earlier,
the dependence of $f_{\rho}^{(2B)}$ on the filling
fraction, which is concentrated in the $\eta^2$ term of
the first line of Eq.~(\ref{eq:f2Blin}), exactly cancels
that of Eq.~(\ref{eq:weak2A}).  Hence, there is some reason to
believe that, when expressed in terms of the channel conductance
$g$, the result is universal in the sense that it is
independent of the details of the
band structure for energies much greater than $U_{2}$,
where $U_{2}$ is much less than the bandwidth $W$.

It is difficult to get a better handle on this
algebraic smorgasbord by mere inspection.
One can add some precision to the picture of what has been accomplished by
first assembling the $g^2$ terms of $f_{\rho}$ and
then plugging in $\rho = 1$ to obtain the contribution to
the symmetric-dot fractional peak splitting $f$.
Upon recalling that
\begin{displaymath}
f^{(2)}_{\rho} = f^{(2A)}_{\rho} + f^{(2B)}_{\rho,(N_{\text{ch}})^2}
                                  + f^{(2B)}_{\rho,N_{\text{ch}}} \, ,
\end{displaymath}
one can evaluate the ${\em A}_{\text{i}}$ integrals numerically for
$\rho = 1$ to obtain
\begin{equation}
f^{(2)}  \approx
           [0.1491]N_{\text{ch}} g^2
           - [0.009798](N_{\text{ch}})^2 g^2 \, .
\label{eq:f2}
\end{equation}
We see that the conjecture about the $(N_{\text{ch}})^2$ and
$N_{\text{ch}}$ contributions to $f^{(2)}$ is correct:
the terms have opposite
sign, and the ratio of their magnitudes is on the order of 10.
For the case of $N_{\text{ch}}=2$, the $g^2$ term
provides the desired upward correction to the $f$-versus-$g$
curve.

Before specializing to the result for $N_{\text{ch}}=2$,
we should explore the consequences of having a term proportional
to $N_{\text{ch}} g^2$.  This term makes the result sensitive to the
``fine structure'' of the interdot conductance.  As remarked
in the introduction, terms of the form $(N_{\text{ch}} g)^{n}$
can be rewritten as a simple power of the total conductance between
the dots: $(N_{\text{ch}} g)^{n} = (g_{\text{tot}})^{n}$.
Should the conductances in the various tunneling channels be
allowed to differ, the form of these terms when written in terms
of $g_{\text{tot}}$ would remain unchanged.  The only alteration
would be in the equation for $g_{\text{tot}}$ itself, which would
revert to the more fundamental form
\begin{equation}
g_{\text{tot}} = \sum_{\sigma} g_{\sigma} \, ,
\label{eq:gtot}
\end{equation}
where $g_{\sigma}$ denotes the dimensionless conductance of the
$\sigma$th channel.

For terms proportional to $(N_{\text{ch}})^{m} g^{n}$
with $ m \neq n$,
the situation is quite different.  Consider the $N_{\text{ch}} g^2$
term in Eq.~(\ref{eq:f2}).  If we had suspended the sum over
channels until the end of our calculation, we would have seen
that these terms are proportional to
\begin{equation}
[ g^2 ]_{\text{tot}} = \sum_{\sigma} g_{\sigma}^2 \, .
\label{eq:gsqrtot}
\end{equation}
Only when symmetry considerations constrain all the
individual channel conductances to be equal
can we safely use
$[ g^2 ]_{\text{tot}} = (g_{\text{tot}})^2/N_{\text{ch}}$.

Consequently, for the general situation in which the
conductances in the separate channels are not necessarily
equal, the appropriate equation for the fractional peak
splitting is the following:
\begin{equation}
f \approx 0.1405 (g_{\text{tot}})
          + 0.1491 [g^2]_{\text{tot}}
          - 0.009798 (g_{\text{tot}})^2 + \ldots
\label{eq:f2gen}
\end{equation}

If we extended the expression to $n$th order in
the dimensionless conductances,
it would contain factors such as
\begin{displaymath}
[g^{m}]_{\text{tot}} = \sum_{\sigma} g_{\sigma}^{m} \, ,
\end{displaymath}
where $m \leq n$ and these factors
appear both alone and in combination up to
$n$th order in dimensionless conductance.
The details of the ``fine structure''
are fully characterized by the set of
$[g^{m}]_{\text{tot}}$
for $ 1 \leq m \leq N_{\text{ch}}$, and
the fractional
peak splitting can be expressed
in terms of these.  Further
modifications might be thought necessary to account
for the ``hyperfine structure'' that
results from allowing the tunneling amplitude $t$
in Eq.~(\ref{eq:weakham}) to be a nontrivial function
of ${\bf k_1}$ and ${\bf k_2}$.  However, as long
as the tunneling amplitude varies little over an
energy range of order $U_{2}$, one would not expect
Eq.~(\ref{eq:f2gen}) to be changed substantially.

\section{The Strong-Coupling Limit for $N_{\text{ch}}=2$}

The $g^2$ correction to the two-channel
solution boosts confidence in the small-$g$ end of our
$f$-versus-$g$ interpolation (see Fig. 3) but does little
to improve the precision of theoretical predictions in
the strong-coupling limit, a fact of particular concern
for the experimentally relevant case of two interdot
tunneling channels.~\cite{Waugh1,Waugh2,Molen}
The sections of the paper that follow improve the
strong-coupling theory for $N_{\text{ch}} = 2$ in two substantial ways.
The first contribution, presented here in Sec. III, is the calculation of
the second term in the $(1-g)$ expansion about the
$g=1$ ground state.  This term, which is linear in $(1-g)$,
is of interest both because it is significant in determining
the shape of the $f$-versus-$g$ curve and because, in the
calculation that yields the primary $(1-g)\ln(1-g)$
term,~\cite{Matveev2} the $(1-g)$ term is naively
ultraviolet-divergent.
The second important contribution, which comes in Sec. IV, is
the provision of powerful evidence that the
coefficients of the leading terms in the $(1-g)$ expansion
are indeed independent of the high-energy
structure of the theory.

To calculate in the limit of $g \rightarrow 1$, we
model the tunneling link between the dots as a
one-dimensional channel with a delta function
scattering potential at its center.  This model was
originally developed for the problem of a single dot
connected to a bulk lead~\cite{Matveev2,Flensberg}
but was shown in Refs.~4 and 5 to be easily
adaptable to that of a pair of coupled dots.
Within this ansatz,
the value of the double-dot charging energy is
a simple reflection of the
total number of electrons that have been transferred
through this channel from one side of the barrier (dot 1)
to the other (dot 2).
In addition, as the system is effectively one-dimensional,
the fermionic degrees of freedom can be bosonized, and
the Euclidean action assumes a characteristic
Luttinger-liquid form~\cite{Matveev2,Flensberg,Kane}:
\begin{eqnarray}
S_{\ \ } & = & \ S_{0} + S_{\text{int}} + S_{\text{b}} \, ,
                         \nonumber \\
S_{0 \ } & = & \ \frac{1}{\beta} \sum_{\sigma} \sum_{\omega_{m}}
     |\omega_{m}|
        |\tilde{\theta}_{\sigma}(\omega_{m})|^{2} \, ,
            \nonumber \\
S_{\text{int}} & = & \
  U_{2} \int_{0}^{\beta} \! d\tau \,
   \left( \frac{1}{\sqrt{\pi}}
     \left[ \sum_{\sigma} \theta_{\sigma} (\tau) \right]
     - \frac{\rho}{2} \right)^{2} \, ,
            \nonumber \\
S_{\text{b} \ } & = & \ \frac{\tilde{V} W}{2 \pi}
   \sum_{\sigma}
   \int_{0}^{\beta} \! d\tau \,
       \cos [2 \sqrt{\pi} \theta_{\sigma} (\tau)] \, .
\label{eq:strongmod1}
\end{eqnarray}
In these formulas,
$\theta_{\sigma}(\tau)$ is a bosonic field that
tracks the displacement of the one-dimensional electron gas
at the barrier ($x=0$), and
$\tilde{\theta}_{\sigma}( \omega_{m})$
is its Fourier transform:
\begin{equation}
\theta_{\sigma}(\tau) = \frac{1}{\beta} \sum_{\omega_{m}}
                     e^{-i \omega_{m} \tau}
                     \tilde{\theta}_{\sigma}( \omega_{m}) \, ,
\label{eq:thetaft}
\end{equation}
where $\tau$ is an imaginary time divided by $\hbar$,
$\beta$ is the inverse temperature
($\beta = 1/\text{k}_{\text{B}}T$), and
$\omega_{m}$ is $\hbar$ times a bosonic Matsubara
frequency ($\omega_{m} = 2 \pi m/\beta$).
In addition, $\tilde{V}$ is a measure of the
barrier strength defined by
$\tilde{V} = \frac{V_{0}}{\hbar v_F}$
for the delta-function potential $V_{0} \delta (x)$.
As for the remaining parameters, $v_F$ is the
Fermi velocity and, as in the weak-coupling theory,
$W$ is the bandwidth---the difference between the lowest
and highest single-particle energies in the channel.
The inverse temperature $\beta$ will
be taken to infinity in calculating the energy
of the ground state.

As in the weak-coupling theory, we ultimately
want to parametrize the coupling between the dots
by the dimensionless channel conductance $g$, rather
than the barrier strength $\tilde{V}$.  Accordingly,
we need to find the relation between $g$ and
$\tilde{V}$.  In our single-mode channel, $g$ equals the
single-particle transmission probability $T$, and
$(1-g)$ equals the reflection probability $R$.
The leading dependence of the channel conductance
on $\tilde{V}$ equals what one would guess from
the reflection probability of a single particle
incident upon a one-dimensional delta-function
potential~\cite{Baym}:
\begin{equation}
(1-g) = \tilde{V}^2 + {\cal O}(\tilde{V}^4) \, .
\label{eq:refamp1}
\end{equation}
Inverting this formula, we find that
\begin{equation}
\tilde{V}^2 = (1-g) + {\cal O}[(1-g)^2] \, .
\label{eq:refamp2}
\end{equation}
To lowest order, we have the approximation of
Matveev,~\cite{Matveev2} $\tilde{V} = \sqrt{1-g}$,
which---it will be seen---is all that is required
for the calculations in this paper.

Having prepared ourselves to switch from a solution
in terms of $\tilde{V}$ to one in terms of $g$,
we proceed with the calculation of the ground-state
energy.  Our first move is to reorganize
the action, expressing it in terms of bosonic fields
that characterize the net charge and pseudospin degrees
of freedom, where the pseudospin degrees of freedom
correspond to ``true spin'' only if
the two channels correspond to spin-up and spin-down,
respectively.
Defining the charge field by
$\theta_{\text{c}} = \theta_1 + \theta_2 + \sqrt{\pi}\rho/2$
and the pseudospin field by
$\theta_{\text{s}} = \theta_1 - \theta_2$,
we find
\begin{eqnarray}
S_{\ \ } & = & \ S_{0}^{\text{(s)}} + S_{0}^{\text{(c)}} + S_{\text{b}}
     \, ,
                    \nonumber \\
S_{0 \ }^{\text{(s)}} & = & \
       \frac{1}{2 \beta} \sum_{\omega_{m}} |\omega_{m}|
         |\tilde{\theta}_{\text{s}}(\omega_{m})|^{2} \, ,
            \nonumber \\
S_{0 \ }^{\text{(c)}} & = & \
       \frac{1}{2 \beta} \sum_{\omega_{m}}
        \left( |\omega_{m}| + \frac{2 U_{2}}{\pi} \right)
        |\tilde{\theta}_{\text{c}}(\omega_{m})|^{2} \, ,
            \nonumber \\
S_{\text{b} \ } & = & \ \frac{\tilde{V} W}{\pi}
   \int_{0}^{\beta} \! d\tau \,
       \cos \! \left[ \sqrt{\pi} \theta_{\text{c}}(\tau)
              + \frac{\pi \rho}{2} \right]
                     \nonumber \\
   & & \mbox{\hspace{0.6in}} \times
      \cos \! \left[ \sqrt{\pi} \theta_{\text{s}}(\tau) \right] \, .
\label{eq:strongmod2}
\end{eqnarray}

The Euclidean action has now been written in terms of
``high-energy'' charge modes and ``low-energy'' pseudospin
modes.  We advance by integrating out the ``high-energy''
charge degrees of freedom.  This integration is
analogous to a renormalization in which one integrates
out the higher-energy degrees of freedom within a
particular channel.~\cite{Fisher}
One begins with the generating functional for the
Euclidean action of Eq.~(\ref{eq:strongmod2}):
\begin{equation}
Z = \int \! D[\theta_{\text{s}}(\tau)] \!
     \int \! D[\theta_{\text{c}}(\tau)] \,
      e^{-S[\theta_{\text{s}}(\tau),
            \theta_{\text{c}}(\tau)]} \, ,
\label{eq:genfnc1}
\end{equation}
where, as usual, time-ordering is implicit within
the functional integral approach.
One then performs the integration over the
fast modes to obtain the generating functional
for an effective action depending
only on the slow modes:
\begin{eqnarray}
Z_{\text{s}} & = & \int \! D[\theta_{\text{s}}(\tau)] \,
          e^{-S_{\text{eff}}[\theta_{\text{s}}(\tau)]} \, ,
                      \nonumber \\
e^{-S_{\text{eff}}[\theta_{\text{s}}(\tau)]}
  & = &  \frac{ e^{-S_{0}^{\text{(s)}} }
           \int \! D[\theta_{\text{c}}(\tau)] \,
             e^{-S_{0}^{\text{(c)}} }
             e^{ - S_{\text{b}} } }
           { \int \! D[\theta_{\text{c}}(\tau)] \,
             e^{-S_{0}^{\text{(c)}} }  } \, .
\label{eq:effact1}
\end{eqnarray}

Eq.~(\ref{eq:effact1}) determines the effective action
$S_{\text{eff}}$.  To solve for it,
one Taylor-expands the exponential factor
$e^{-S_{\text{b}}}$, performs the integral over
charge degrees of freedom, and re-exponentiates
the result.  Before doing any of this, it is
useful to make the following definition:
\begin{equation}
      \langle \hat{A} \rangle_{\text{c}}
       = \frac{\int \!  D[\theta_{\text{c}}(\tau)] \,
                  \hat{A} \, e^{-S_{0}^{\text{(c)}}} }
              {\int \!  D[\theta_{\text{c}}(\tau)] \,
                   e^{-S_{0}^{\text{(c)}}} } \, .
\label{eq:chargeave}
\end{equation}
One can then rewrite
Eq.~(\ref{eq:effact1}) as follows:
\begin{eqnarray}
e^{-S_{\text{eff}}}
  & = & e^{-S_{0}^{\text{(s)}} } \langle e^{-S_{\text{b}} }
              \rangle_{\text{c}}
                            \nonumber \\
  & = & e^{-S_{0}^{\text{(s)}}}
         \left[ 1 - \langle S_{\text{b}} \rangle_{\text{c}}
                + \frac{1}{2}
                \langle S_{\text{b}}^{2} \rangle_{\text{c}}
                + {\cal O}(\tilde{V}^3) \right] \, .
\label{eq:effact2}
\end{eqnarray}
Upon re-exponentiation, one obtains
\begin{eqnarray}
S_{\text{eff}}
   & = & S_{0}^{\text{(s)}}
            + \left\langle S_{\text{b}} \right\rangle_{\text{c}}
          -\frac{1}{2} \left\langle \left[ S_{\text{b}}
            - \left\langle S_{\text{b}} \right\rangle_{\text{c}}
                                    \right]^{2}
                                          \right\rangle_{\text{c}}
          + {\cal O}(\tilde{V}^3) \, .
\label{eq:effact3}
\end{eqnarray}
It is clear  that to solve for the effective action to order
\mbox{$\tilde{V}^{2} = (1-g)$},
we must solve for both corrections to $S_{0}^{\text{(s)}}$
on the right side of Eq.~(\ref{eq:effact3}).

Details of the calculation of these terms are presented
in Appendix~B.  The result is that
\begin{displaymath}
S_{\text{eff}}  =
      S_{0}^{\text{(s)}} + S_{\text{b}}^{(1)} + S_{\text{b}}^{(2)} \, ,
\end{displaymath}
where
\begin{eqnarray}
S_{\text{b}}^{(1)}
    & = & \frac{\tilde{V} W}{\pi} e^{-\frac{\pi}{2} K_{\text{c}}(0)}
          \cos \! \left(\frac{\pi \rho}{2} \right)
          \int_{0}^{\beta} \! d\tau \,
      \cos \! \left[ \sqrt{\pi} \theta_{\text{s}}(\tau) \right] \, ,
                                \nonumber \\
S_{\text{b}}^{(2)}
   & = & \left( \frac{\tilde{V} W}{\pi} \right)^{2}
           e^{- \pi K_{\text{c}}(0)}
          \int_{0}^{\beta} \! d\tau_1 \! \int_{0}^{\tau_1} \! d\tau_2 \,
                                \nonumber \\
   &   & \mbox{\hspace{0.3in}} \times \{
            \cos^2 \! \left( \frac{\pi \rho}{2} \right)
               \left[ 1 - e^{-\pi K_{\text{c}}(\tau_1 - \tau_2)} \right]
                                    \nonumber \\
   &   & \mbox{\hspace{0.6in}}
               - \sinh[\pi K_{\text{c}}(\tau_1 - \tau_2)] \}
                                 \nonumber \\
   &   & \mbox{\hspace{0.3in}} \times
         \cos \! \left[ \sqrt{\pi} \theta_{\text{s}}(\tau_1) \right]
         \cos \! \left[ \sqrt{\pi} \theta_{\text{s}}(\tau_2) \right] \, .
\label{eq:moments}
\end{eqnarray}
The function $K_{\text{c}}(\tau)$ that appears in these formulas
is the charge-channel correlation function,
$K_{\text{c}}(\tau) = \langle \theta_{\text{c}}(\tau)
                              \theta_{\text{c}}(0) \rangle_{\text{c}}$.
Its numerical value can be found from the formula
\begin{equation}
K_{\text{c}}(\tau) = \frac{1}{\pi} \text{Re}
    \int_{0}^{\infty} \! d\omega \,
\frac{e^{-\left( \frac{2}{W} +{\em i}\tau \right)\omega}}
               {\omega + \frac{2 U_{2}}{\pi}} \, .
\label{eq:Kcorrel}
\end{equation}

To progress further, we define a new ``unperturbed action''
$S_{\text{New}} = S_{0}^{\text{(s)}} + S_{\text{b}}^{(1)}$.
We then write down the Hamiltonian that corresponds to this
action:
\begin{eqnarray}
H_{\text{New}} & = & H_{0}^{\text{(s)}} + H_{\text{b}}^{(1)} \, ,
                             \nonumber \\
H_{\text{b}}^{(1)} & = &
          \frac{\tilde{V} W}{\pi} e^{-\frac{\pi}{2} K_{\text{c}}(0)}
          \cos \! \left(\frac{\pi \rho}{2} \right)
      \cos \! \left[ \sqrt{\pi} \theta_{\text{s}}(0) \right] \, .
\label{eq:Hnew}
\end{eqnarray}
This is the Hamiltonian diagonalized by Matveev in
Ref.~9 through a process of
``debosonization'' (see Appendix~B) in which the
Hamiltonian is rewritten in terms of fermion operators
$d_k$ and $d$:
\begin{eqnarray}
H_{0}^{\text{(s)}} & = &
     \int_{-\Lambda}^{\Lambda} \! dk \,
                    \xi_{k} d_{k}^{\dagger} d_{k} \, ,
                                \nonumber \\
H_{\text{b}}^{(1)} & = &   \lambda
    \int_{-\Lambda}^{\Lambda} \! dk \,
     \left[ d_{k}^{\dagger} (d + d^{\dagger})
           + (d + d^{\dagger}) d_{k} \right] \, .
\label{eq:Hnewferm}
\end{eqnarray}
Here the single-particle energy $\xi_{k}$, the fermion
interaction parameter $\lambda$, and the wave-vector
cut-off $\Lambda$ have the formulas
$\xi_{k} = \hbar v_F k$,
$\lambda = \tilde{V} \cos (\pi \rho/2)
                 \sqrt{2 e^\gamma \hbar v_F U_{2}/\pi^3}$,
and $\Lambda = W/2\hbar v_F$.

Since the Hamiltonian is now quadratic in fermion operators,
a Bogoliubov transformation brings it to the desired diagonal
form:
\begin{equation}
H_{\text{New}} =
  E_{\text{New}}^{(0)} + \int_{0}^{\Lambda} \! dk \, \xi_{k}
        \left( {C}_{k}^{\dagger} {C}_{k}
              +\tilde{C}_{k}^{\dagger} \tilde{C}_{k} \right) \, ,
\label{eq:Hnewdiag}
\end{equation}
where, if we write down only the terms of lowest order in
$\tilde{V}$, replacing all others by an ellipsis,
\begin{eqnarray}
\tilde{C}_{k} & = & \frac{1}{\sqrt{2}}(d_{k} + d_{-k}^{\dagger}) \, ,
                          \nonumber \\
{C}_{k} & = & \frac{1}{\sqrt{2}}(d_{k} - d_{-k}^{\dagger}) + \ldots
\label{eq:bogops}
\end{eqnarray}

The correction to the $\tilde{V}=0$ ground-state energy
is produced by the omitted terms in $C_{k}$
(for details, see Appendix~B).  In particular,
using $\Delta_{\text{str}}^{(1)}$ to represent the
difference between the ground-state energies of
$H_{\text{New}}$ for $\tilde{V}=0$ and for arbitrary
$\tilde{V}$, respectively, one finds that
\begin{eqnarray}
\Delta_{\text{str}}^{(1)} (\rho) & = &
     \frac{4 e^{\gamma} U_{2}}{\pi^3} \tilde{V}^2
            \cos^2 \! \left( \frac{\pi \rho}{2} \right)
                    \nonumber \\
      & & \mbox{\hspace{0.3in}} \times
          \left( \ln \! \left[ {\tilde{V}^2
                  \cos^2 \! \left( \frac{\pi \rho}{2} \right)} \right]
  - \ln \! \left[ \frac{\psi}{2} \right] \right.
                    \nonumber \\
      & & \hspace{0.6in} \left.
  - 1 + \ln \! \left[ \frac{8 e^{\gamma}}{\pi^2} \right]  \right) \, .
\label{eq:Del1str}
\end{eqnarray}
As before, $\psi = W/U_{2}$, where the bandwidth $W \gg U_{2}$.

We see that the result for this first correction
contains terms that are quadratic in $\tilde{V}$ and
logarithmically divergent in $\psi$.  This ultraviolet
divergence is circumvented in Ref.~9 by
the statement that one should replace $W$ by $U_{2}$
because keeping only the first
term from charge-channel integration is only a good approximation
for energies less than the charging energy $U_{2}$.
The terms in Eq.~(\ref{eq:Del1str}) that are merely quadratic in
$\tilde{V}$ are thereby rendered finite and can be dropped
in favor of the leading
$\tilde{V}^2\cos^2 (\pi \rho/2) \ln[\tilde{V}^2\cos^2 (\pi \rho/2)]$
dependence.

To eliminate the logarithmic divergence more formally, one must
calculate the shift in the ground-state energy that is induced by
$S_{\text{b}}^{(2)}$ [recall Eq.~(\ref{eq:moments})].
As this term is itself quadratic in $\tilde{V}$
and as we are only interested in knowing the ground-state
energy to order $\tilde{V}^2$, we can drop all but the
leading part of the $S_{\text{b}}^{(2)}$-induced shift.
In expressing $S_{\text{b}}^{(2)}$ in terms of the
diagonalizing operators of $H_{\text{New}}$,
one may use the truncated formulas of Eq.~(\ref{eq:bogops}).
The relevant shift in the ground-state energy is then found
by calculating the expectation value of $S_{\text{b}}^{(2)}$ in the
ground state of $H_{\text{New}}$ (see Appendix~B):
\begin{eqnarray}
\Delta^{(2)}_{\text{str}}(\rho) & = &
   \frac{4 e^{\gamma} U_{2}}{\pi^3} \tilde{V}^2
            \cos^2 \! \left( \frac{\pi \rho}{2} \right)
                         \nonumber \\
 & & \mbox{}      \times
            \int_{0}^{\infty} dx \,
\left[ 1 - e^{-\pi K_{\text{c}} \left( \frac{2x}{W} \right)} \right]
  \frac{1 - e^{-x}}{x} \, ,
\label{eq:Del2str}
\end{eqnarray}
where units have been chosen such that $\hbar = 1$ and terms
independent of $\rho$ have been dropped since they are not relevant
to evaluation of the fractional peak splitting $f$.
It is not too hard to see that the
factor $[1 - e^{-\pi K_{\text{c}}(2x/W)}]$ in the integrand
makes for an ultraviolet cut-off of order $\psi = W/U_{2}$
(see Appendix~B).  It is even easier to see that
$(1-e^{-x})$ provides an infrared
cut-off of order $1$.  Thus, one can surmise that the leading
term from the integral is $\ln (\psi/2)$, which is precisely
what is needed to cancel the ultraviolet divergence in
$\Delta^{(1)}_{\text{str}}$.

What remains is to calculate the rest of the
integral in Eq.~\ref{eq:Del2str}, which we call $\Phi$:
\begin{eqnarray}
\Phi & = & \lim_{ \psi \rightarrow \infty} \left(
             \int_{0}^{\infty} dx \,
\left[ 1 - e^{-\pi K_{\text{c}} \left( \frac{2x}{W} \right)} \right]
  \frac{1 - e^{-x}}{x} \right.
                         \nonumber \\
  & & \mbox{\hspace{0.6in}} \left. -\ln \! \left[ \frac{\psi}{2} \right]
\right)
\, .
\label{eq:Phi}
\end{eqnarray}
Numerical approximation of the integral in the limit
$\psi \rightarrow \infty$ gives
$\Phi = 0.1703 \pm 0.0002$.

One can now sum $\Delta^{(1)}_{\text{str}}$
and $\Delta^{(2)}_{\text{str}}$
to get the strong-coupling energy shift through order
$\tilde{V}^2$.  Having dropped terms that are independent
of $\rho$, one has
\begin{eqnarray}
\Delta_{\text{str}}(\rho) & = &
     \frac{4 e^{\gamma} U_{2}}{\pi^3} \tilde{V}^2
            \cos^2 \! \left( \frac{\pi \rho}{2} \right)
 \left( \ln \!
\left[ \tilde{V}^2 \cos^2 \!\left(\frac{\pi \rho}{2} \right) \right]
   \right.
                       \nonumber \\
  & & \mbox{\hspace{0.3in}} \left.
-1 + \Phi + \ln \left[\frac{8 e^{\gamma}}{\pi^2} \right]  \right) + \ldots
\label{eq:Delstrongtot}
\end{eqnarray}

We can now straightforwardly compute the fractional peak-splitting
$f$ in terms of the dimensionless conductance $g$.
As mentioned before, if we are only interested in obtaining
the ground-state energy to order $(1-g)$,
only the leading term of Eq.~(\ref{eq:refamp2}) is relevant
in converting Eq.~(\ref{eq:Delstrongtot}) to an expression
in terms of $(1-g)$.  The value of $f=f_{\rho=1}$
follows from the fact that, in the strong-coupling limit,
\begin{equation}
f_{\rho} = 1- \frac{\Delta_{\text{str}}(\rho)
                   -\Delta_{\text{str}}(0)}{U_{2} \rho^2/4} \, .
\label{eq:frhostr}
\end{equation}
In particular, Eqs.~(\ref{eq:refamp2}), (\ref{eq:Delstrongtot}),
and (\ref{eq:frhostr}) yield
\begin{eqnarray}
f  & = & 1 + \frac{16 e^{\gamma}}{\pi^3} (1-g)\ln (1-g)
                      \nonumber \\
  & & \mbox{} - \frac{16 e^{\gamma}}{\pi^3}
\left[1 -\ln \! \left(\frac{8 e^{\gamma}}{\pi^2} \right) - \Phi \right](1-g)
   + \ldots
\label{eq:fstr1}
\end{eqnarray}

\noindent Since $\Phi \approx 0.1703$, we have
\begin{equation}
f \approx 1 + 0.919 (1-g) \ln(1-g) - 0.425 (1-g) + \ldots
\label{eq:fstr2}
\end{equation}

Having determined the first corrections to the leading behaviors
for both $g \rightarrow 0$ and $g \rightarrow 1$,
we now have a more plausible picture for the connection between
the $N_{\text{ch}}=2$ weak- and strong-coupling limits (see Fig.~3).
The fit to the data could be improved if the
interdot capacitance were larger than experimentally
estimated~\cite{Waugh2,Golden1} or if asymmetry between the
dots were important.~\cite{Matveev3,Matveev4}
In any case, whether or not such further emendations should
be made, the theory is within the range of present
experimental error.  The corrections introduced in this
paper have moved the weak- and strong-coupling
predictions by reasonable amounts
in the desired directions, increasing both the ease and
the precision of interpolation between
the weak- and strong-coupling limits.

\begin{minipage}{3.27truein}
  \begin{figure}[H]
    \begin{center}
      \leavevmode
      \epsfxsize=3.27truein
      \epsfbox{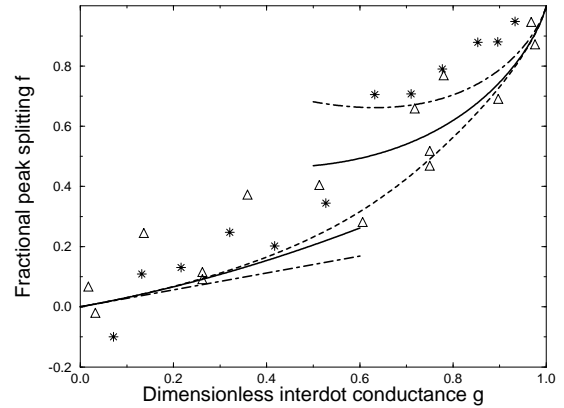}
    \end{center}
    \caption{Graph of the fractional Coulomb blockade conductance
peak splitting $f$ as
a function of the dimensionless conductance per channel $g$
in the weak- and strong-tunneling limits for
\mbox{$N_{\text{ch}}=2$}.  The new theoretical curves
are depicted as solid lines.  The old theoretical curves
from Refs.~4 and 5 are
dot-dashed lines.
%The strong-coupling curves are
%obtained from Eq.~(19) under the assumption that $C_{2}=1$.
The dashed curve shows a possible interpolating function.
Data points from Refs.~2 and 3 are given as
triangles or stars; the two different symbols correspond to
different data sets.
The value of $f$ for the experimental data has been extracted
from the measured splitting fraction $f'$ by using the
method discussed in Ref.~4 with experimentally estimated
values of $20 \mbox{ }$aF for the constant
interdot capacitance and $0.4 \mbox{ }$fF for the
total single-dot capacitance.$^{2,3}$}
    \label{fig:fofg}
  \end{figure}
  \smallskip
\end{minipage}

\section{Insensitivity to the High-Energy Density of States}

\subsection{Insensitivity to Functional Form of Bosonic Cut-Off}

To have confidence that our coupled-dot calculations can be
usefully compared to empirical data, we should make sure
that the result, $f$ expressed as a function of $g$,
is independent of the details of the band structure far
from the Fermi surface, where the assumption of a constant
density of states becomes invalid.  We have done much
to confirm such robustness in the regime of weak coupling,
for we have shown there that, through second order in $g$,
$f(g)$ is independent of the bandwidth $W$ and the
filling fraction $F$ as long
as both $FW$ and $(1-F)W$ are much larger than the charging
energy $U_{2}$.  Such dual invariance
indicates that we can simply shear off a
nontrivial number of high-energy states without affecting
the result.  We would expect then that we could make less
Draconian modifications of the high-energy density of states
with similarly perfect impunity.

With regard to the strong-coupling theory, matters have
been left less assured.  In Ref.~4,
we introduced a factor of $C_2$ multiplying the first
term in Eq.~(\ref{eq:fstr1}) to guard against the possibility
that the coefficient of the energy shift calculated in the
Luttinger-liquid approach was partly a product of the
approach itself and, in particular,
the manner in which the ultraviolet cut-off
was imposed.  Concern about such a possibility arises from
the fact that the leading term in the $(1-g)$ expansion is
proportional to the product of
$e^{-\frac{\pi}{2} K_{\text{c}}(0)}$ and ${\cal A}$,
where ${\cal A}$ is the generalization to non-exponential
cut-offs of the normalization
factor in Eq.~(\ref{eq:fermibose}) that gives the
proportionality between the fermionic position operators and
the exponentials of bosonic fields:
\begin{equation}
\psi_{f}^{\dagger}(0,\tau) = {\cal A} e^{i \sqrt{\pi} \phi_{f}(\tau)} \, .
\label{eq:defA}
\end{equation}
Changing the nature of the bosonic cut-off
[e.g., from the exponential $e^{-\alpha |\omega|}$
to the Gaussian $e^{-(\pi/4)\alpha^2 \omega^2}$ ]
causes the value of
$e^{-\frac{\pi}{2} K_{\text{c}}(0)}$ to be multiplied by a
constant factor.  Although one would hope that a similar
shift in the value of ${\cal A}$ compensates for the
change in $e^{-\frac{\pi}{2} K_{\text{c}}(0)}$, to the
authors knowledge, such a happy circumstance has not
previously been checked to be true.

Similar questions could be asked about the prefactor for
the term linear in $(1-g)$, with which we associate a
factor $C_{3}$, where $C_{3} = 1$ for the Luttinger-liquid
approach with the standard exponential cut-off.
This term is proportional both to
$|{\cal A}|^2 e^{- \pi K_{\text{c}}(0)}$ and to an integral
that depends upon $e^{-\pi K_{\text{c}}(\tau)}$ [see
Eq.~(\ref{eq:Del2str}) in Sec. III and
Eq.~(\ref{eq:shft2C}) in Appendix B].
Hence, in order to prove that the two leading strong-coupling
terms do not vary with the choice of cut-off
function, one must show that neither
the product ${\cal D}_1 = |{\cal A}|^2 e^{- \pi K_{\text{c}}(0)}$
nor the integral
\begin{eqnarray}
{\cal D}_2 & = & \frac{1}{\beta W} \int_{0}^{\beta W/2} \! dx \,
  \left( \frac{\beta W}{2} - x \right)
                 \nonumber \\
 & & \mbox{\hspace{0.6in}} \times
  \left[ 1 - e^{-\pi K_{\text{c}}(2x/W)} \right]
  \frac{1 - e^{-x}}{x}
\label{eq:integC3}
\end{eqnarray}
assumes different values when the shape of the cut-off is
changed.
Though we do not have a general proof that ${\cal D}_1$
and ${\cal D}_2$
are independent of the cut-off function, we can show that
they remain the same for a whole class of functions that
includes the exponential cut-off and that they are similarly
unchanged when one replaces the exponential cut-off by
a Gaussian.  We believe that these facts are convincing
evidence that the prefactors in Eq.~(\ref{eq:fstr1}) are
insensitive to the nature of the high-energy cut-off.

First, we prove that ${\cal D}_1$ and ${\cal D}_2$
are the same for all cut-offs of the form
\begin{equation}
\nu(\omega,\alpha,\{b_{m}\}) =
e^{- \alpha |\omega|}
  \left( 1 + \sum_{m=1}^{M} b_{m} \alpha^{m} |\omega|^{m} \right) ,
\label{eq:seriescut}
\end{equation}
where either $M$ is finite or, for large $m$,
$b_{m}$ falls to zero faster than $m^{-\zeta}/m!$
for some real $\zeta > 0$.  As usual, it
is assumed that $\alpha U_{2} \ll 1$, where $\alpha = 2/W$.
We add the further assumption that
$\left[ (m-1)! \, b_{m} \alpha U_{2} \right] \ll 1$
for all $m$.

The first step in our proof is to solve
for the change in $e^{-\pi K_{\text{c}}(0)}$ when one goes
from the standard exponential cut-off $\nu(\omega,\alpha,\{0\})$ to
the more general form $\nu(\omega,\alpha,\{b_{m}\})$.
The formula for $K_{\text{c}}(\tau)$ [recall
Eq.~(\ref{eq:Kcorrel})] becomes
\begin{equation}
K_{\text{c}}(\tau) = \frac{1}{\pi} \text{Re}
    \int_{0}^{\infty} \! d\omega \, \, \nu(\omega, \alpha,\{b_{m}\})
     \, \frac{ e^{-i \tau \omega} }
               {\omega + \frac{2 U_{2}}{\pi}} \, .
\label{eq:Kcgen}
\end{equation}
We can write the change in $K_{\text{c}}(0)$ as
\begin{equation}
\delta \! K_{\text{c}}(0) = \frac{1}{\pi}
     \sum_{m=1}^{M} b_{m} \int_{0}^{\infty} \! d\omega
        \frac{ \alpha^{m} \omega^{m} e^{-\alpha \omega} }
             {\omega + \frac{2 U_{2}}{\pi} }.
\label{eq:deltKc1}
\end{equation}
Using $\omega = (\omega + 2 U_{2}/\pi - 2 U_{2}/\pi)$ and
the binomial theorem, we can expand $\omega^{m}$ in powers
of $(\omega + 2 U_{2}/\pi)$.  The integration is then
straightforward and yields
\begin{equation}
\pi \delta \! K_{\text{c}}(0) =
     \sum_{m=1}^{M} b_{m} (m-1)! \, [1 + {\cal O}(\alpha U_{2})].
\label{eq:deltKc2}
\end{equation}
Dropping the correction, we have the result
\begin{equation}
e^{-\pi K_{\text{c}}(0)} =
     \left[ e^{-\sum b_{m} (m-1)!} \right]
     \, e^{-\pi K_{\text{c}, 0}(0) } \, ,
\label{eq:seriesKc}
\end{equation}
where $K_{\text{c}, 0}(\tau)$ is the correlation function
for the standard exponential cut-off.

Calculation of the change in the normalization constant
${\cal A}$ is more complicated.
Following V.~J.~Emery,~\cite{Emery} we find
\begin{equation}
|{\cal A}|^{-2} =
   \int_{-\infty}^{\infty} \! dx \,
   e^{ \int_{0}^{\infty} \! d\omega \, \nu(\omega, \alpha, \{b_{m}\}) \,
     \frac{e^{i \omega x/\hbar v_F} - 1 }{\omega} }
      + c.c.
\label{eq:Agen1}
\end{equation}
A bit of calculation reveals that
\begin{eqnarray}
|{\cal A}|^{-2} & = &
   \left[ e^{-\sum_{m=1}^{M} b_{m} (m-1)! \, } \right]
       \int_{-\infty}^{\infty} \! dx \,
      \left( \frac{\alpha}{\alpha - i x} \right)
                \nonumber \\
 & & \mbox{\hspace{0.6in}} \times
       e^{ \sum b_{m} (m-1)! \,
            \left( \frac{\alpha}{\alpha - i x} \right)^{m} }
   + c.c.
\label{eq:Aseries1}
\end{eqnarray}

It is apparent that the bracketed factor
exactly cancels the factor that
multiplies $e^{-\pi K_{\text{c}, 0}(0)}$ in
Eq.~(\ref{eq:seriesKc}).  Thus, in order for
${\cal D}_1 = |{\cal A}|^2 e^{- \pi K_{\text{c}}(0)}$
to be unaltered, the value of the integral in
Eq.~(\ref{eq:Aseries1}) cannot change as the
$b_{m}$ are varied.  In short, the partial derivative of
the integral with respect to each of these coefficients
must be zero.  The partial derivative with respect to $b_{m}$ is
given by the following formula:
\begin{eqnarray}
P_{m} & = &  (m-1)! \, \int_{-\infty}^{\infty} \! dx
      \left( \frac{\alpha}{\alpha - i x} \right)^{m+1}
                      \nonumber \\
 & &  \mbox{\hspace{0.3in}} \times
 e^{ \sum b_{m} (m-1)! \,
            \left( \frac{\alpha}{\alpha - i x} \right)^{m} }
   + c.c.
\label{eq:Apart1}
\end{eqnarray}
Let $z = \alpha/(\alpha - i x)$.  The resulting integral
in the complex $z$ plane follows a closed path, beginning and
ending at $z = 0$:
\begin{equation}
P_{m} =  -i \alpha (m-1)! \, \oint \! dz \, z^{m-1} \,
      e^{ \sum b_{m} (m-1)! \, z^{m} }
    + c.c.
\label{eq:Apart2}
\end{equation}
For $M$ finite or $b_{m}$ falling off faster than
$m^{-\zeta}/m!$,
the integrand is analytic throughout the region enclosed by
the contour.  By Cauchy's Theorem, $P_{m} = 0$.

Since the integral of Eq.~(\ref{eq:Aseries1}) does not
vary with $b_{m}$, we can make the statement
\begin{equation}
|{\cal A}|^{2} =
   \left[ e^{\sum_{m=1}^{M} b_{m} (m-1)! \,} \right]
   |{\cal A}_{0}|^{2},
\label{eq:Aseries2}
\end{equation}
where ${\cal A}_{0}$ is the normalization factor for
the standard exponential cut-off.
Eqs.~(\ref{eq:seriesKc}) and (\ref{eq:Aseries2}) yield
\begin{equation}
{\cal D}_{1} = |{\cal A}|^{2} e^{-\pi K_{\text{c}}(0)} =
     |{\cal A}_{0}|^{2}
     e^{-\pi K_{\text{c}, 0}(0)}.
\label{eq:seriesC2}
\end{equation}
We have now shown that, for the class of cut-offs
$\nu(\omega,\alpha,\{b_{m}\})$, $C_{2}$ is
constant.

What about $C_{3}$?  To determine its fate, we must
find the change in the quantity ${\cal D}_{2}$
[recall Eq.~(\ref{eq:integC3})].  After substituting
$\alpha$ for $(2/W)$, we follow essentially
the same path that we blazed in determining the change in
$K_{\text{c}}(0)$ and find that the change in
$K_{\text{c}}(\alpha x)$ is given by the formula
\begin{equation}
\delta \! K_{\text{c}}(\alpha x) =
   \frac{1}{2 \pi} \sum_{m=1}^{M} b_{m} (m-1)! \,
      \left[ \frac{1}{(1 + i x)^{m} }
                + c.c. \right] .
%                + \frac{1}{(1 - i x)^{m} } \right].
\label{eq:deltKcx}
\end{equation}
Employing this, we can break the integral
on the right side of Eq.~(\ref{eq:integC3}) into
two parts, the first of which is from $0$ to $\psi^{1-\epsilon}$,
where $0 < \epsilon < 1$ and $\psi = (W/U_{2}) \gg 1$.
In this interval, the contribution from the
entire term proportional to $e^{-\pi K_{\text{c}}(\alpha x)}$
can be shown to be zero in the limit $\psi \rightarrow \infty$.
In the remainder,
$\delta \! K_{\text{c}}(\alpha x)$ is on the order of
$1/x^2$, which implies that the correction due to the
generalization of $\nu(\omega,\alpha,\{b_{m}\})$
is proportional to
\begin{displaymath}
\int_{\psi^{1-\epsilon}}^{\beta W/2} \! dx \,
     \frac{e^{-\pi K_{\text{c}, 0}(2x/W)} } {x^3}
\leq \int_{\psi^{1-\epsilon}}^{\beta W/2} \frac{dx}{x^3} \, ,
\end{displaymath}
which also equals zero in the limit $\psi \rightarrow \infty$.
Therefore, ${\cal D}_{2}$ is constant, and we have
proven that our strong-coupling results are insensitive
to varying the cut-offs within the class
$\nu(\omega, \alpha, \{b_{m}\})$.

The values of ${\cal D}_{1}$ and ${\cal D}_{2}$ can
be shown to be similarly unaltered when we switch
from the exponential cut-off to a Gaussian:
\begin{equation}
\nu_{G}(\omega,\alpha)
   = e^{-\frac{\pi}{4} \alpha^2 \omega^2}.
\label{eq:Gausscut}
\end{equation}
Solving for $e^{-\pi K_{\text{c}}(0)}$ with this
weight function, one discovers that
\begin{equation}
e^{-\pi K_{\text{c},G}(0)}
   = \frac{e^{\gamma/2}}{\sqrt{\pi}} \alpha U_{2},
\label{eq:GuassKc}
\end{equation}
where $\gamma$ is once again the Euler-Mascheroni
constant.  The normalization coefficient ${\cal A}_G$
has not been solved for analytically.  However,
starting from Eq.~(\ref{eq:Agen1}), one finds
that
\begin{eqnarray}
|{\cal A}_{G}|^{-2} & = & \pi \alpha \int_{0}^{\infty} \! dx
      \cos \left[ \frac{\pi}{2} \text{Erf}(x/2) \right]
              \nonumber \\
 & & \mbox{\hspace{0.6in}} \times
  e^{ \frac{i \sqrt{\pi}}{2} \int_{0}^{x} \! dy \,
           e^{-y^{2}/4} \, \text{Erf}(i y/2) },
\label{eq:Agauss}
\end{eqnarray}
where $\text{Erf}(x) = (2/\sqrt{\pi}) \int_{0}^{x} \! dt \, e^{-t^2}$
is the error function.  It has been confirmed numerically
that through at least 12 digits the product
${\cal D}_{1, G}
= |{\cal A}_{G}|^{2} e^{-\pi K_{\text{c},G}(0)}$
agrees with the exponential cut-off.
By arguments similar to those used for the class
of cut-off functions studied above, it has also been
shown that in the limit $\psi \rightarrow \infty$,
the integral ${\cal D}_{2, G}$ is the same
as for the exponential.  The coefficients in Eq.~(\ref{eq:fstr1})
are again unaltered,
and it seems reasonable to suppose that the
invariance is general.

\subsection{Insensitivity to Fermionic Filling Fraction}

Thus, it appears fairly certain that modifying the
high-energy density of states in the bosonized
theory does not affect the results of Sec.~III.
Nevertheless, having solved the weak-coupling model for
the general case of a fermionic system not necessarily
at half-filling and having seen that, when expressed as
functions of the tunneling amplitude, both the conductance
and the fractional peak splitting depend
upon the filling fraction, one might wonder what
happens to the strong-coupling results
when one begins with a fermionic system that is not
necessarily at half-filling.  Since
Luttinger-style bosonization assumes symmetry between
occupied and empty states,
such a system can only be properly bosonized
after the asymmetric fermion states have been integrated
out.  For example, if the system is below half-filling
[$F < (1-F)$] and the zero of energy is at the Fermi surface
($\epsilon_F = 0$), the fermionic single-particle states
with energies between $FW$ and $(1-F)W$ must be integrated
out, leaving a symmetric effective theory with
single-particle energies ranging from $-FW$ to $FW$.  Only
after this symmetrization can the theory be bosonized
without losing knowledge of the fermionic filling fraction
$F$.

The task before us, therefore, is to ``symmetrize'' the
fermionic theory that lies behind the bosonized action
of Eq.~(\ref{eq:strongmod1}).  The archetypal fermionic
Hamiltonian consists of the usual three parts:
the single-particle kinetic energies,
the multiparticle potential energy, and the backscattering
barrier.  The Hamiltonian therefore takes the following form:
\begin{eqnarray}
H_{\ } & = & H_{K} + H_{C} + H_{B} \, , \nonumber \\
H_{K} & = & \sum_{j=1}^{2} \sum_{\sigma} \sum_{k}
     \xi_{k} c^{\dagger}_{j k \sigma}
                             c_{j k \sigma} \, ,
      \nonumber  \\
H_{C} & = & U_{2} (\hat{n} - \rho/2)^{2} \, ,
                        \nonumber  \\
H_{B}  & = & \sum_{\sigma} \sum_{k_1 k_2}
             v(c^{\dagger}_{2 \bf{k_2} \sigma}
           c_{1 \bf{k_1} \sigma} + \mbox{H.c.}) \, ,
\label{eq:strmodferm}
\end{eqnarray}
where $\xi_{k} = \hbar v_F k$ and $j$ is the index that
distinguishes between right-movers ($j = 1$)
and left-movers ($j = 2$).

The operator $\hat{n}$ is now somewhat more complicated
than in the weak-coupling theory.
In its simplest form, it can be written as
\begin{equation}
\hat{n} = \frac{1}{2} \sum_{j=1} \int \! dx \,
             [ \Theta(x) - \Theta(-x) ] \,
             \psi_{j}^{\dagger}(x) \psi_{j}(x) \, ,
\label{eq:nstr1}
\end{equation}
where $\Theta(x)$ is the Heaviside step function and
$\psi_{j}$ is the annihilation operator in position space
for a right-moving ($j = 1$) or a left-moving ($j = 2$)
fermion.
After writing the components of the integrand in the momentum
representation and integrating over $x$, one finds that,
for a one-dimensional system of length $L$,
\begin{equation}
\hat{n} = \frac{-i}{L} \sum_{j} \sum_{\sigma} \sum_{k_1 k_2}
        \frac{c^{\dagger}_{j k_2 \sigma} c_{j k_1 \sigma}}
             {k_2 - k_1} (1 - \delta_{k_1, k_2}) \, ,
\label{eq:nstr2}
\end{equation}
which is equivalent to the integral version obtained by
Matveev~{\cite{Matveev1}} from the observation
that $d\hat{n}/dt$ equals the current operator
at $x=0$, the point of ``division'' between the two dots.
(This point is, of course, not entirely well-defined
in the limit $g \rightarrow 1$.)

The above equations for the Hamiltonian and number operator
are presented as discrete sums.
For future reference in implementing the symmetrization of
the theory, we write the components of our Hamiltonian
in integral form:
\begin{eqnarray}
H_{K} & = & \left( \frac{\hbar v_F}{\delta} \right)
             \sum_{j=1}^{2} \int \! dk \, \xi_k \,
                    c^{\dagger}_{j k \sigma}
                             c_{j k \sigma} \, ,
      \nonumber  \\
H_{C} & = & U_{2} (\hat{n} - \rho/2)^{2} \, ,
                        \nonumber  \\
H_{B}  & = & v \left( \frac{\hbar v_F}{\delta} \right)
               \sum_{\sigma} \int \! dk_1 \! \int \! dk_2 \,
               (c^{\dagger}_{2 \bf{k_2} \sigma}
            c_{1 \bf{k_1} \sigma} + \mbox{H.c.}) \,
\label{eq:strmodint}
\end{eqnarray}
where $\delta$ is the level-spacing for the one-dimensional
system ($\delta = 2\pi \hbar v_F/L$) and
\begin{equation}
\hat{n} =
      \frac{-i}{2\pi} \sum_{j} \sum_{\sigma}
      {\cal P} \int \! dk_1 \! \int \! dk_2
        \frac{c^{\dagger}_{j k_2 \sigma} c_{j k_1 \sigma}}
             {k_2 - k_1} \, .
\label{eq:nstr3}
\end{equation}

For the fermionic strong-coupling model of
Eq.~(\ref{eq:strmodferm}), calculation of
the channel conductance between the dots proceeds along the same
lines as for weak coupling (see Ref.~4).
In fact, since the density of states is constant in both
theories, setting $U_{2} = 0$---the first step in the conductance
calculation---renders them essentially identical, the only
differences being in the last term, where $t$ has been
replaced by $v$ and the index $i$ for dot-1 or
dot-2 fermions has been replaced by the index $j$ for
right-movers or left-movers.  Accordingly, unlike the
weak-tunneling term $H_{T}$,
the perturbation $H_{B}$ scatters fermions backward instead of
transporting them forward and therefore causes a reduction
in the conductance of the unperturbed system.  Recalling the
size of the conductance induced by $H_{T}$ in the weak-coupling
model, it is not hard to see that the channel conductance in
the strong-coupling model is given by
\begin{equation}
g = 1 - \frac{4 \chi}{|1 + (1+i\eta)^2 \chi|^2} \, \, ,
\label{eq:gstrferm}
\end{equation}
where $\chi = (\pi v/\delta)^2$ and
$\eta = (1/\pi)\ln[F/(1-F)]$.
As in the weak-coupling theory,
the result becomes troublesome as $\chi$ becomes large.
However, we should be able to trust its testimony that
the filling fraction does not affect the interdot
conductance through second-order
in $(\pi v/\delta)$.

This is all we need to know, for $(\pi v/\delta)$ can be
straightforwardly written in terms of our previous
strong-coupling parameter  $\tilde{V}$.  The relation
is $\tilde{V} = 2(\pi v/\delta)$, and it follows that
$\chi = (\tilde{V}/2)^2$.
%It is not hard to see that $v = V_{0}/L$, where $L$ is the
%length of the one-dimensional system.  From the fact
%that $\delta = 2\pi \hbar v_F/L$, it follows that
%$\tilde{V} = 2\pi v/\delta$ and $\chi = \tilde{V}/2$.
We recover the leading-order result of Eq.~(\ref{eq:refamp2})
and see that, to order $\tilde{V}^2$, the
channel conductance $g$ is independent of the
fermionic filling fraction $F$.
If we can likewise show that the relation between
$\tilde{V}$ and the differential energy shift
\begin{equation}
\delta \! \Delta(\rho) =
     [\Delta_{\text{str}}(\rho) - \Delta_{\text{str}}(0)]
\label{eq:deldel}
\end{equation}
does not depend on the filling fraction $F$,
we will know that the same is true for our final
strong-coupling result,
the expression for $f(g)$ in Eq.~(\ref{eq:fstr1}).

To prove $\delta \! \Delta$'s invariance with respect
to $F$, we symmetrize the fermionic theory through
a renormalization in which we integrate out all
single-particle states at an energy distance of
$W'/2$ or more from the Fermi surface, where
$U_2 \ll W' \ll W$.  The resulting symmetric
theory with bandwidth $W'$ can be
bosonized without further qualm.  However,
as renormalization generates terms that are not
present in the original Hamiltonian, we must check
to see what relevant effects these have upon the
low-energy theory.  We must also keep track of
any contributions to $\delta \! \Delta$
that arise from the high-energy degrees of freedom alone.

Before we go about doing this, a comment on our
approach is in order.  One might view the proposed
renormalization as occurring in two distinct stages:
first, we integrate out the asymmetric particle-hole
states; then, we integrate both particle and
hole states down to energy $W'$.  Since all that
we will need to consider are the general scaling
properties of the terms generated during the
renormalization process, the distinction between
the stages is of no importance and is henceforth
ignored.

The argument resumes.
Since our interest is in the Coulomb blockade, the
renormalization scheme we use is designed to leave the
Coulombic interaction term $H_{C}$ unchanged.
After wave-vectors between the original wave-vector
cut-off $\Lambda$ and the new wave-vector cut-off
$\Lambda/b$ (where $b > 1$)
have been integrated out, the theory is re-scaled by
writing it in terms of a new set of wave-vectors
$k_{b} = b k$.
Invariance of $H_{C}$ is achieved by re-scaling the
fermion creation and annihilation operators as well:
$c^{\dagger}_{j k_{b} \sigma} = b^{-1/2} c^{\dagger}_{j k \sigma}$.
[One might prefer to say that the coherent-state
Grassman variables that correspond to the
operators are re-scaled (see Ref.~38).]
The effect of renormalization upon the parameters
$(\hbar v_F/\delta)$, $U_{2}$, and $v(\hbar v_F/\delta)$ of
Eq.~(\ref{eq:strmodferm}) is as follows:
\begin{eqnarray}
\left[ \frac{\hbar v_F}{\delta} \right]' & = &
      b^{-1} \left[ \frac{\hbar v_{F}}{\delta} \right] \, , \nonumber \\
\left[U_{2}\right]' & = & U_{2} \, , \nonumber \\
\left[v(\hbar v_F/\delta)\right]'
   & = & b^{-1} \left[v(\hbar v_F/\delta)\right] \, .
\label{eq:renorm1}
\end{eqnarray}
The backscattering $H_{B}$ is revealed to be
dangerously irrelevant.  Though
it scales like an irrelevant term, we cannot safely set
it to zero as we know from Eq.~(\ref{eq:Del1str})
that the energy shift is singular as $v \rightarrow 0$.

In addition to rescaling the terms in the original
Hamiltonian, renormalization generates terms of its
own.  It is not hard to see that all but the new backscattering
terms are irrelevant.  The original
Hamiltonian consists of the kinetic energy $H_{K}$, a two-body
interaction $H_{C2}$, a one-body interaction
$H_{C1}$, and a backscattering term $H_{B}$.  $H_{C2}$ and
$H_{C1}$ are normal-ordered operators given by the following
formulas:
\begin{eqnarray}
H_{C2} & = & \frac{-U_{2}}{(2\pi)^2} \sum_{j_1, j_2}
   \sum_{\sigma_1, \sigma_2}
   {\cal P} \int \! dk_1 \ldots dk_4 \,
          \nonumber \\
  & & \mbox{\hspace{0.6in}} \times
   \frac{c^{\dagger}_{j_2 k_4 \sigma_2} c^{\dagger}_{j_1 k_2 \sigma_1}
          c_{j_1 k_1 \sigma_1} c_{j_2 k_3 \sigma_2} }
        {(k_4 - k_3)(k_2 - k_1)} \, \, , \nonumber \\
H_{C1} & = & -\rho U_{2} \hat{n} +
\frac{-U_{2}}{(2\pi)^2} \sum_{j} \sum_{\sigma}
   {\cal P} \int \! dk_1 \! \int \! dk_2 \,
     \frac{c^{\dagger}_{j k_2 \sigma} c_{j k_1 \sigma} }
        {k_2 - k_1} \,
              \nonumber \\
 & & \mbox{\hspace{0.6in}} \times
   \left( \ln\left|\frac{\Lambda - k_1}{\Lambda + k_1}\right|
         - \ln\left|\frac{\Lambda - k_2}{\Lambda + k_2}\right| \right)
                          \, ,
\label{eq:HC1&2}
\end{eqnarray}
where $\Lambda = W/\hbar v_F$.
$H_{C2}$, $H_{C1}$, and $H_{B}$ can be represented by Feynman graphs
[see Fig. 4(a)], which can then be connected to construct the
terms that renormalization adds to the Hamiltonian.
As usual, the internal lines of the second-generation graphs
carry only high-energy momenta which lie within the shell of
wave-vectors that are integrated out.

\begin{minipage}{3.27truein}
  \begin{figure}[H]
    \begin{center}
      \leavevmode
      \epsfxsize=3.27truein
      \epsfbox{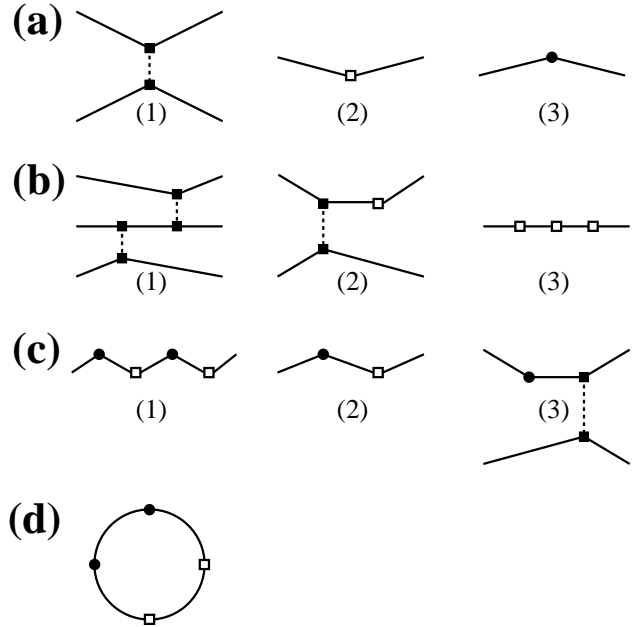}
    \end{center}
    \caption{Feynman diagrams for integrating out single-particle
energies a distance greater than $W'$ from the Fermi surface in
the fermionic version of the strong-coupling theory.
(a) The three building-block Feynman graphs.
Diagram 1 corresponds to the two-body Coulomb interaction
$H_{C2}$. Diagram 2 corresponds to the one-body Coulomb term
$H_{C1}$.  Diagram 3 represents the backscattering $H_{B}$.
(b) Second-generation $m$-body graphs constructed by contracting
$H_{C2}$'s and $H_{C1}$'s.  These terms are all irrelevant to
the low-energy theory, scaling to zero under renormalization.
(c) Second-generation graphs formed from combinations of
$H_{B}$, $H_{C2}$, and $H_{C1}$.  Terms such as Diagram 1
that contain an even number of $H_{B}$'s are irrelevant under
renormalization.
Diagrams 2 and 3 involve odd numbers of $H_{B}$'s and are
therefore dangerously irrelevant.
Nevertheless, they
are negligible in size compared to corresponding low-energy
graphs and therefore can be safely discarded.
(d) An example of a closed diagram used to calculate the
contribution to the energy shift from the degrees of
freedom that correspond to single-particle energies
more than $W'$ from the Fermi surface.}
    \label{fig:feyndiags}
  \end{figure}
  \smallskip
\end{minipage}

Given such rules for making second-generation terms,
one can deduce that, whenever one creates a new
term by connecting
lines emanating from the $H_{C2}$ and $H_{C1}$ graphs
[see Fig. 4(b) for examples], one
picks up a scaling factor of $b^{-1}$.
For example, Diagram 2 of Fig. 4(b) represents a
two-body interaction produced by contracting one
$H_{C2}$ with one $H_{C1}$.  This new interaction
term is similar to $H_{C2}$ except that the denominator
contains only one power of $(k_4 - k_3)$ or $(k_2 - k_1)$
and, consequently, is less singular than $H_{C2}$,
which is fixed under rescaling.  Thus, the second-generation
term must shrink under renormalization.
%Such a link results from a contraction of two powers of
%$\hat{n}$ such as the following:
%\begin{equation}
%\hat{n}---\hat{n} =
%        \frac{-i}{2\pi} \sum_{s_1 s_2}
%              \sum_{\sigma_1 \sigma_2} \int dk_1 \ldots dk_4
%        \frac{c^{\dagger}_{s_2 k_4 \sigma_2}
%    \bar{ c_{s_2 k_3 \sigma_2} c^{\dagger}_{s_1 k_2 \sigma_1} }
%              c_{s_1 k_1 \sigma_1} }
%             {k_4 - k_3}{k_2 - k_1} \, ,
%\label{eq:ncontrct}
%\end{equation}
%where the two operators beneath the bar are contracted.
%The contraction forces $k_3 = k_2 \equiv k$.  The momentum
%denominator can then be rewritten as
%\begin{equation}
%\frac{1}{(k_4 - k)(k - k_1)} =
%   \frac{1}{k_4 - k_1}
%   \left( \frac{1}{k - k_1} + \frac{k_4 - k} \right).
%\label{eq:ndenom1}
%\end{equation}
Indeed, all such graphs formed from contracting
the Coulombic interaction terms are similarly
irrelevant and scale to zero under renormalization.
They can be ignored in
the effective theory.  We should expect this result.
Otherwise, our Coulomb blockade model would
probably never have been useful at all.

As for graphs that involve the backscattering term
$H_{B}$ [see Fig. 4(c)], we need
only consider these to order $v^2$, for we go no further
in calculating $f(g)$.  Depending upon how many
Coulombic interaction terms are introduced, the
second-generation graphs that contain $H_{B}$
all scale down by at least
a factor of $b^{-1}$.  Consequently, all but those which
contribute to low-energy backscattering are irrelevant.
Thus, we can drop graphs such as Diagram 1 of Fig.~4(c) that
contain an even number of $H_{B}$'s.  Graphs containing an
odd number of $H_{B}$'s are dangerously irrelevant but
can ultimately be ignored because
they are negligible compared to the corresponding graphs
that can be constructed from the low-energy portions
of the original $H_{C2}$, $H_{C1}$, and $H_{B}$.
Diagrams 2 and 3 of Fig. 4(c), for example, are of order
$v(U_{2}/W')$.  If we had renormalized down to
$W''$, where $U_{2} \ll W'' \ll W'$, we would have found
the corresponding graphs to be
of order $v(U_{2}/W'')$.  The contribution from energies
above $W'$ is therefore seen to be merely perturbative
in relation to the contribution from energies between
$W''$ and $W'$.  The conclusion is that we can drop the
parts of the graphs produced by integrating over energies
greater than $W'$.  Returning to our original renormalization
down to $W'$, we see that the graphs produced here have been
shown to be negligible.  The argument that the symmetrizing
renormalization does not cause any significant changes
in the low-energy Hamiltonian is complete.

Having disposed of the concern that the process of
symmetrization might leave us with important
new low-energy terms,
we now show that any constant terms produced
are similarly insignificant.  Such
constant terms correspond to closed diagrams constructed from
the original Feynman graphs.  Since all lines are internal,
they all represent the propagation of high-energy excitations.
There are obviously an infinite number of closed diagrams.
Fortunately, we can limit our attention to a certain
subset.  We need not concern
ourselves with diagrams involving less than two
$H_{C1}$ graphs: diagrams with only one $H_{C1}$
graph must sum to zero as $\Delta(\rho)$ is even
in $\rho$; diagrams with zero $H_{C1}$ graphs
cannot contribute to the differential energy shift
$\delta \! \Delta$.
Similarly, in any pertinent closed graph,
$H_{B}$ must appear a nonzero and
even number of times.  It cannot be absent as
terms that do not include it
shift all relevant ground-state energies
equally and are therefore unimportant.
Furthermore, in any closed graph, it must appear an
even number of times because $H_{B}$ is the only term that exchanges
right- and left-movers.
Thus, all the diagrams we need consider consist of
a nonzero and even number of $H_{B}$'s, at least
two $H_{C1}$'s, and an arbitrary number of $H_{C2}$'s
[see Fig.~4(d) for a canonical example].

Each such diagram
corresponds to a number of time-ordered terms in
Rayleigh-Schr\"{o}dinger perturbation theory.  For a Feynman
diagram with $r$ internal lines, the associated
Rayleigh-Schr\"{o}dinger terms have $r$ integrations
over momenta and $(r-1)$ propagators with denominators
linear in the momenta.  If the $H_{C2}$ graph appears
$m_2$ times in the Feynman diagram and the $H_{C1}$
graph appears $m_1$ times, there are
$m = (2 m_2 + m_1) \geq 2$ additional denominators
linear in the momenta, which have their origin in the
wave-vector denominator of $\hat{n}$ [recall
Eqs.~(\ref{eq:nstr3}) and (\ref{eq:HC1&2})].
The propagator
denominators are always on the order of $W'$
or greater.  The $\hat{n}$ denominators
are of the form $(k - k')$, where $k$ and $k'$ are both
in the high-energy wave-vector shell.  Thus, these
denominators can go to
zero.  However, the contribution from the
regions where they become zero is negligible, the
somewhat simplified explanation being that,
when one of them goes to zero, the rest of the integrand
can be treated as essentially constant, and we have
\begin{displaymath}
{\cal P} \int_{-\epsilon \Lambda'}^{\epsilon \Lambda'}
     \frac{1}{k}
\left[1 + {\cal O}\left(\frac{\hbar v_F k}{W'}\right)\right]
= {\cal O}(\epsilon),
\end{displaymath}
where $\Lambda' = W'/\hbar v_F$ and the constant
$\epsilon \ll 1$.
It follows that contributions to the overall result
only come when the $\hat{n}$ denominators are themselves
of order $W'/\hbar v_F$.

As a result, what remains is a nonsingular integration over
$r$ momenta of an integrand that is proportional to
$[\bar{k}_1 \ldots \bar{k}_{(r-1+m)}]^{-1}$, where
the $\bar{k}_i$ are linear in the momenta over which
we integrate.  Noting that the only other momentum
dependence comes from the logarithmic term of
$H_{C1}$, we see that, in energy units, the result of
the integration is of order $(1/W')^{m-1}$.
We now multiply the
result of our integration by the various factors
of $U_{2}$, $v$, and $\delta$ that stand aside the
integral.  For a closed diagram in which $H_{B}$,
$H_{C2}$, and $H_{C1}$ appear $j$, $m_2$,
and $m_1$ times, respectively,
the contribution to the energy shift is readily
seen to be of order
$U_{2} (v/\delta)^{2 j} (U_{2}/W')^{m-1}$, where
$m = (2 m_2 + m_1) \geq 2$.  As $(U_{2}/W') \ll 1$ and
the overall energy shift is of order $U_{2}$,
these terms are negligible.

Thus, at least to order $v^2$, integrating out
all particle and hole
excitations at distances greater than $W'/2$ from
the Fermi surface produces neither relevant new
terms in the low-energy Hamiltonian nor significant
constant contributions to the differential energy shift.
As what remains is a fermionic theory at half-filling,
the result for $f(g)$ in Sec.~III is unaffected
by possible ``high-energy'' deviations from this condition,
an important property if we wish to compare our
predictions with empirical data.
We would hope that a similarly universal solution
for $f(g)$ could be found to higher orders in $(1-g)$.
However, if the formula for
the interdot conductance [recall Eq.~(\ref{eq:gstrferm})]
is correct to some non-leading order, such overall
independence of the filling fraction must---as  in
the weak-coupling limit---come through cancellation
of the separate filling-fraction dependences of the conductance
and the energy shift when one is expressed in terms
of the other.  If this were shown to be true, we would
see once again that the interdot conductance $g$ and
not the bare matrix element for tunneling or reflection
is the correct parameter to achieve a universal
description of the coupling dependence of a double-dot
Coulomb blockade.

\section{Conclusion}

The present paper substantially improves the results of earlier
theoretical work on the Coulomb-blockade peak-splitting for two
coupled quantum dots~\cite{Golden1,Matveev3,Matveev4} making an
important contribution to the growing body of theoretical and
experimental work on such coupled-dot
systems~\cite{Waugh1,Waugh2,Molen,Ruzin,Ford,Stafford1,%
Stafford2,Kemerink,Tsukada,Sakamoto,Klimeck,Hofmann,%
Haug,Vaart,Crouch,Pals}.
By extending the weak-coupling theory to second order in $g$
for arbitrary $N_{\text{ch}}$, it has shown how the small-$N_{\text{ch}}$
theory crosses over to the large-$N_{\text{ch}}$ theory in the
neighborhood of $N_{\text{ch}} \approx 10$.  Furthermore, it
has demonstrated that, at least for the leading two terms  in the
weak-coupling theory, the channel conductance $g$ is the
``correct'' parameter to use in constructing a theory for
the peak splitting that is universal in the sense that it is
does not depend on the high-energy band structure.
Finally, this paper has made the $N_{\text{ch}}=2$ theory both
stronger and broader---broader in that the sub-leading term
is calculated; stronger in that the leading and sub-leading
terms for strong-coupling are confirmed to be insensitive to
the manner in which the high-energy cut-off is taken.
Thus, the paper has made more plausible efforts
to connect weak- and strong-coupling behaviors and to compare
theoretical results with the data from recent two-channel
experiments.~\cite{Waugh1,Waugh2,Molen}

\section*{Acknowledgments}
The authors are grateful for helpful conversations with
F. R. Waugh, R. M. Westervelt, C. H. Crouch, C. Livermore,
A. L. Moustakas, S. H. Simon, and S. Ramanathan.
J. M. G. thanks the United States Air Force for financial
support.  This work was also supported by the NSF
through the Harvard Materials Research Science and
Engineering Center, Grant No. DMR94-00396.

\end{multicols}

\widetext

\appendix
\section{Details of the Weak-Coupling Calculation}

As described in Sec. II, the procedure in evaluating the
fourth-order energy shift is to calculate the
$(N_{\text{ch}})^2$ and $N_{\text{ch}}$ terms
separately.  Calculation of the $(N_{\text{ch}})^2$ terms
is facilitated by rewriting them in terms of two
energy variables instead of four.  Calculation of the
$N_{\text{ch}}$ terms is made easier by differentiating
twice with respect to $\rho$ while performing the
integrations over energy and then integrating
twice with respect to $\rho$ at
the end.  Terms that are constant or linear with
respect to $\rho$ cancel in the final result, the
relative energy shift $(\Delta_{0} - \Delta_{\rho})$,
so we have not lost useful information as a result of the
double differentiation.

As mentioned in Sec. II, the ``wrinkle'' in these
computations, the appearance of integrals of the form
\begin{displaymath}
{\cal P} \int_{0}^{R \psi} dx \, \, \frac{\ln (x + B)}{x + A} \ ,
\end{displaymath}
is resolved by Taylor-expanding the logarithm about
$(x+A)$ for $(B-A)<|x+A|$ and about $(B-A)$ for
$(B-A)>|x+A|$.  For $A < 0$, one first breaks the integral
into the intervals $(0,|A|-\epsilon)$ and $(|A|+\epsilon,R\psi)$.
After this, one proceeds as usual.  The results are
\begin{eqnarray}
{\cal P} \int_{0}^{R \psi} dx \, \frac{\ln (x + B)}{x + A}
  & = &  \frac{1}{2} \ln^2(R\psi+A) - \frac{1}{2} \ln^2 A \nonumber \\
  &   & \mbox{ \ \ \ \ } + \sum_{n=1}^{\infty}
           \frac{(-1)^{n+1}}{n^2}
           \left[   \left(\frac{B-A}{A} \right)^{n}
                 - \left(\frac{B-A}{R\psi+A} \right)^{n} \right]
         \nonumber \\
  &   & \mbox{ \hspace{1in} \text{for} \ } 0 < (B-A) < A < (R\psi+A) ,
         \nonumber \\
  &   & \mbox{ \hspace{1in} } \nonumber \\
  & = & \frac{1}{2} \ln^2 (R\psi+A) - \ln (B-A) \ln A
        +\frac{1}{2} \ln^2 (B-A)  +  \frac{\pi^2}{6} \nonumber \\
  &   & \mbox{ \ \ \ \ }
       - \sum_{n=1}^{\infty} \frac{(-1)^{n+1}}{n^2}
          \left[   \left(\frac{A}{B-A} \right)^{n}
                 + \left(\frac{B-A}{R\psi+A} \right)^{n} \right]
         \nonumber \\
  &   & \mbox{ \hspace{1in} \text{for} \ } 0 < A < (B-A) < (R\psi+A) ,
         \nonumber \\
  &   & \mbox{ \hspace{1in} } \nonumber \\
  & = &  \ln (B-A) [ \ln (R\psi+A) - \ln A ] \nonumber \\
  &   & \mbox{ \ \ \ \ } + \sum_{n=1}^{\infty}
           \frac{(-1)^{n+1}}{n^2}
           \left[   \left(\frac{R\psi+A}{B-A} \right)^{n}
                 - \left(\frac{A}{B-A} \right)^{n} \right]
         \nonumber \\
  &   & \mbox{ \hspace{1in} \text{for} \ } 0 < A < (R\psi+A) < (B-A) ,
         \nonumber \\
  &   & \mbox{\hspace{1in}} \nonumber \\
  & = &  \frac{1}{2}\ln^2(R\psi-|A|)
         -\ln(B+|A|)\ln|A| + \frac{1}{2}\ln^2(B + |A|) +\frac{\pi^2}{6}
         \nonumber \\
  &   & \mbox{ \ \ \ \ }
         + \sum_{n=1}^{\infty} \frac{1}{n^2}
                    \left( \frac{|A|}{B+|A|} \right)^{n}
         - \sum_{n=1}^{\infty}
                \frac{(-1)^{n+1}}{n^2}
                \left( \frac{B+|A|}{R\psi-|A|} \right)^{n}
         \nonumber \\
  &   & \mbox{ \hspace{1in} \text{for} \ } A < 0 < (B+|A|) < (R\psi-|A|) ,
         \nonumber \\
  &   & \mbox{ \hspace{1in} }
         \nonumber \\
  & = & \ln(B+|A|)[\ln(R\psi-|A|) - \ln|A|]
         \nonumber \\
  &   & +\sum_{n=1}^{\infty} \frac{1}{n^2}
            \left( \frac{|A|}{B+|A|} \right)^{n}
        +\sum_{n=1}^{\infty}
            \frac{(-1)^{n+1}}{n^2}
            \left( \frac{R\psi-|A|}{B+|A|} \right)^{n}
         \nonumber \\
  &   & \mbox{ \hspace{1in} \text{for} \ } A < 0 < (R\psi-|A|) < (B+|A|) .
\label{eq:logint}
\end{eqnarray}
These five integrals are all we need.  In confirming that
the solution for $(B-A)<(R\psi+A)$ evolves continuously into
that for $(R\psi+A)<(B-A)$, it is useful to
recognize~\cite{Spiegel} that
\begin{equation}
\sum_{n=1}^{\infty} \frac{(-1)^{n+1}}{n^2} = \frac{\pi^2}{12}.
\label{eq:pisum}
\end{equation}

Having equipped ourselves to smooth the ``wrinkles,''
we can proceed with a fuller description of calculation
of the fourth-order terms.  The $(N_{\text{ch}})^2$
calculation is reviewed first.  An illustrative
segment of the $N_{\text{ch}}$ calculation follows.

In Sec. II, it was remarked that each of the
$(N_{\text{ch}})^2$ terms could be written in terms of two
energy variables ($\epsilon_I = \epsilon_3 - \epsilon_1$,
$\epsilon_4 - \epsilon_2$) instead of the original four.
The ``cost'' of this conversion is the appearance of
a nontrivial density of states:
\begin{equation}
\int_{0}^{\epsilon_F} \! d\epsilon_1 \!
\int_{0}^{\epsilon_F} \! d\epsilon_2 \!
\int_{\epsilon_F}^{W} \! d\epsilon_3 \!
\int_{\epsilon_F}^{W} \! d\epsilon_4 \,
   h(\epsilon_3 - \epsilon_1, \epsilon_4 - \epsilon_2)
= \int_{0}^{\epsilon_F} \! d\epsilon_{\text{I}} \,
              \nu(\epsilon_{\text{I}}) \!
\int_{0}^{\epsilon_F} \! d\epsilon_{\text{II}} \,
              \nu(\epsilon_{\text{II}}) \,
   h(\epsilon_{\text{I}}, \epsilon_{\text{II}}),
\label{eq:fourtotwo}
\end{equation}
where $\nu(\epsilon)$ is the density of states.
For a system at or below half-filling,
\begin{equation}
\nu(\epsilon) = \left\{ \begin{array}{ll}
   \epsilon & \text{for} \ 0 \leq \epsilon < \epsilon_F, \\
   \epsilon_F & \text{for} \
           \epsilon_F \leq \epsilon < (W-\epsilon_F), \\
   (W - \epsilon) \hspace{0.2in} & \text{for} \
           (W-\epsilon_F) \leq \epsilon < W.
              \end{array} \right.
\end{equation}
(We need not worry about a system above half-filling
as such as system can be mapped to one
below half-filling through an exchange of particles
and holes.)

Using the new variables $\epsilon_{\text{I}}$ and
$\epsilon_{\text{II}}$,
we can sum the integrands for the $(N_{\text{ch}})^2$
terms shown in Fig. 2 (the others
are obtained by taking $\rho \rightarrow -\rho$).
If drop the common factor $-N_{\text{ch}}(t/\delta)^4 U_{2}$,
the result is the following:
\begin{eqnarray}
h_{\text{tot}} & = & \frac{-2}
   {[\epsilon_{\text{I}}+U_2(1-\rho)]^2 [\epsilon_{\text{II}} + U_2(1-\rho)]
     [\epsilon_{\text{II}}+\epsilon_{\text{I}} + U_2(4-2\rho)]}
           \nonumber \\
 &  & \mbox{}
  + \frac{2}{[\epsilon_{\text{I}} + U_2(1-\rho)]^2
    [\epsilon_{\text{II}} + U_2(1+\rho)]
    [\epsilon_{\text{II}} + \epsilon_{\text{I}}]}.
\end{eqnarray}

It is not hard to show that
\begin{displaymath}
\int_{\epsilon_F}^{W} \! d\epsilon_{\text{I}} \!
\int_{\epsilon_F}^{W} \! d\epsilon_{\text{II}} \,
[h_{\text{tot}}(\epsilon_{\text{I}}, \epsilon_{\text{II}},\rho)
 - h_{\text{tot}}(\epsilon_{\text{I}},\epsilon_{\text{II}},0)] = 0
\end{displaymath}
in the limit $\psi = W/U_2 \rightarrow \infty$.
Accordingly, we need only calculate
\begin{displaymath}
\int_{0}^{\epsilon_F} \! d\epsilon_{\text{I}} \!
\int_{0}^{\epsilon_F} \! d\epsilon_{\text{II}} \,
h_{\text{tot}}(\epsilon_{\text{I}}, \epsilon_{\text{II}},\rho).
\end{displaymath}
The process of evaluating this double integral is lengthy but
straightforward.  The only ``wrinkles'' that appear---integrals
of the form of Eq.~\ref{eq:logint}---are no longer problematic.
The end result is Eq.~(\ref{eq:f2Bsqr}) of Sec. II.

We now move to consideration of the fourth-order terms
linear in the number of conducting channels.
Recall that the $(N_{\text{ch}})^2$ terms were added before
the integrations over energy were performed.  This order of
tasks is reversed for the $N_{\text{ch}}$ terms, the computation
of which revolves primarily about finding a favorable
permutation of the operations of differentiating
and integrating with respect to $\rho$, integrating with
respect to the $i$th energy variable, and integrating by parts.
Consequently, perhaps the best way to describe the derivation
of the $N_{\text{ch}}$ contribution is to walk through the
computation of a single illustrative term.
After seeing the methodology employed in calculating this
term, the tireless reader should have little difficulty in
computing the rest.

The representative term we choose is that corresponding to
Diagram~2 of Fig.~2(b).
This term involves an exchange of a pair of electrons and,
consequently, picks up an ``exchange'' minus sign.
The diagram depicts the following sequence of events:
\newcounter{bean}
\begin{list}
   {\Roman{bean}.}{\usecounter{bean}
       \setlength{\rightmargin}{\leftmargin}
       \setlength{\itemsep}{0.0in}
       \setlength{\parskip}{0.0in}
}
  \item Electron 1 tunnels from dot 1 to dot 2, going from a
   single-particle state with kinetic energy $\epsilon_{1}$ to
   one with kinetic energy $\epsilon_{3}$.  The energy of the
   resulting double-dot state relative to that of the
   unperturbed ground state is
   \mbox{$[\epsilon_{3} - \epsilon_{1} + U_{2}(1-\rho)]$}.
  \item Electron 2 tunnels from dot 1 to dot 2, going from a
   single-particle state with kinetic energy $\epsilon_{2}$ to
   one with kinetic energy $\epsilon_{4}$.  The system's energy
   relative to the unperturbed ground state is now
   \mbox{$[\epsilon_{4} + \epsilon_{3} - \epsilon_{2}
             - \epsilon_{1} + 2 U_{2} (2-\rho)]$}.
  \item Electron 2 tunnels back to dot 1, settling into the
   initial single-particle state of Electron 1.
   The ensuing relative system energy is
   \mbox{$[\epsilon_{3} - \epsilon_{2} + U_{2} (1-\rho)]$}.
  \item Electron 1 tunnels back to dot 1, settling into
   Electron 2's initial single-particle state.  The
   unperturbed ground state has been recovered.
\end{list}

With all the intermediate-state energies known, it is easy
to write down the contribution to the fourth-order energy
shift:
\begin{eqnarray}
\Delta^{(4)}_{N_{\text{ch}},2}(\rho) & = & t^4 \sum_{\sigma}
    \sum_{\epsilon_1, \epsilon_2 }
    \sum_{\epsilon_3, \epsilon_4 }
    \frac{1}{[\epsilon_3 - \epsilon_2 + U_{2}(1-\rho)]}
          \nonumber \\
    &   & \mbox{\hspace{0.3in}} \times
         \frac{1}{[\epsilon_{4} + \epsilon_{3} - \epsilon_{2}
             - \epsilon_{1} + 2 U_{2} (2-\rho)]} \times
         \frac{1}{
             [\epsilon_{3} - \epsilon_{1} + U_{2}(1-\rho)]} \, .
\label{eq:exshft1}
\end{eqnarray}
The sums over $\epsilon_1$ and $\epsilon_2$ extend from $0$
to the Fermi energy $\epsilon_F$.  Those for $\epsilon_3$
and $\epsilon_4$ go from $\epsilon_F$ to the bandwidth $W$.
The sum over the channel index $\sigma$ results
from the fact that Electrons 1 and 2 can share any one
of the $N_{\text{ch}}$ tunneling channels.
Though the formula contains such unphysical terms
as that for which $\epsilon_1 = \epsilon_2$, such terms
are down by factors of the level spacing $\delta$ divided
by $\delta/F W$ or $\delta/(1-F)W$, and their inclusion
has no effect in the limit $W/\delta \rightarrow \infty$.

Accordingly, we can cease worrying about these terms, for
we assume that \mbox{$\delta \ll U_{2} \ll W$}, a
postulate that permits us to work in the continuum
limit, replacing the sums in Eq.~(\ref{eq:exshft1})
by integrals:
\begin{eqnarray}
\Delta^{(4)}_{N_{\text{ch}},2}(\rho) & = &
    N_{\text{ch}} \left( \frac{t}{\delta} \right)^4
    \int_{0}^{\epsilon_F} d\epsilon_1 \!
    \int_{0}^{\epsilon_F} d\epsilon_2 \!
    \int_{\epsilon_F}^{W} d\epsilon_3 \!
    \int_{\epsilon_F}^{W} d\epsilon_4 \,
    \frac{1}{[\epsilon_3 - \epsilon_2 + U_{2}(1-\rho)]}
           \nonumber \\
    &  & \mbox{\hspace{0.3in}} \times
        \frac{1}{[\epsilon_{4} + \epsilon_{3} - \epsilon_{2}
             - \epsilon_{1} + 2 U_{2} (2-\rho)]} \times
        \frac{1}{[\epsilon_{3} - \epsilon_{1} + U_{2}(1-\rho)]} \, .
\label{eq:exshft2}
\end{eqnarray}
These integrals can be rewritten in terms of
dimensionless variables $x_i$:
\begin{equation}
x_i  = \left\{ \begin{array}{ll}
                  \frac{\epsilon_F - \epsilon_i}{U_{2}}
                   &  \mbox{for \ } i = 1 \: \mbox{or} \: 2  \\
      \frac{\epsilon_i - \epsilon_F}{U_{2}}
                   &  \mbox{for \ }
                                       i = 3 \: \mbox{or} \: 4 \, .
               \end{array} \right.
\end{equation}
With this choice of integration variables, it becomes clear that
$\Delta^{(4)}_{N_{\text{ch}},2}(\rho)$ is linear
in $U_{2}$.  Specifically, we find that
\begin{eqnarray}
\Delta^{(4)}_{N_{\text{ch}},2}(\rho) & = &
     N_{\text{ch}} \left( \frac{t}{\delta} \right)^4 U_{2} \times
        I(\rho,F,\psi) \, ,  \nonumber \\
I(\rho,F,\psi) & = &
    \int_{0}^{F \psi} \! dx_1  \!
    \int_{0}^{F \psi} \! dx_2   \!
    \int_{0}^{(1-F)\psi} \! dx_3 \!
    \int_{0}^{(1-F)\psi} \! dx_4 \,
    \frac{1}{[x_3 + x_2 + 1-\rho]}
           \nonumber \\
    &  & \mbox{\hspace{0.3in}} \times \frac{1}{[x_4 + x_3 + x_2
             + x_1 + 2 (2-\rho)]} \times
        \frac{1}{[x_3 + x_1 + 1-\rho]} \, .
\label{eq:exshft3}
\end{eqnarray}

All the shuffling of notatation still leaves us
confronted with a quadruple integral.
Opting to postpone a frontal assault, we
try a sidestepping movement, computing
the partial derivative with respect to $\rho$:
\begin{equation}
I_{\rho}  =
         \int_{0}^{F \psi} \! dx_1 \!
         \int_{0}^{F \psi} \! dx_2 \!
         \int_{0}^{(1-F)\psi} \! dx_3 \!
         \int_{0}^{(1-F)\psi} \! dx_4   \left(
         \frac{1}{[ \mbox{ \ } ]^2 [ \mbox{ \ } ] [ \mbox{ \ } ] }
       + \frac{2}{[ \mbox{ \ }] [ \mbox{ \ }]^2 [ \mbox{ \ }]}
       + \frac{1}{[ \mbox{ \ }] [ \mbox{ \ }] [ \mbox{ \ }]^2} \right) \, ,
\label{eq:1deriv}
\end{equation}
where
the subscript $\rho$ signifies that $I_{\rho}$ is
the partial derivative of $I$ with respect to $\rho$
and the brackets on the right-hand side of the equation
have the same contents in the same order as those in
Eq.~(\ref{eq:exshft3}).  As the third term in the
integrand of Eq.~(\ref{eq:1deriv}) differs from the
first only by an exchange of the indices 1 and 2,
we can drop the third term and double the first.
When the enhanced first term is integrated by parts
with respect to $x_2$, the products are two
triple-integral terms and a quadruple-integral term
that exactly cancels the second
term of Eq.~(\ref{eq:1deriv}).  With the definitions
$A_0 = 0$ and $A_1 = F\psi$, we have
\begin{eqnarray}
I_{\rho}  & = & 2
         \sum_{p=0}^{1} (-1)^{p}
         \int_{0}^{F \psi} \! dx_1 \!
         \int_{0}^{(1-F)\psi} \! dx_3 \!
         \int_{0}^{(1-F)\psi} \! dx_4 \,
         \frac{1}{[x_3 + A_{p} + 1-\rho]}
             \nonumber \\
        & & \mbox{\hspace{0.3in}}
         \times \frac{1}{[x_4 + x_3 + x_1 + A_{p} + 2(2-\rho)]}
         \times  \frac{1}{[x_3 + x_1 + 1-\rho]} \, .
\label{eq:2deriv}
\end{eqnarray}

Having benefited once from differentiation with respect to
$\rho$, we try it again.  The second derivative of $I$
with respect to $\rho$ has the following form:
\begin{equation}
I_{\rho \rho}   =  2
         \int_{0}^{F \psi} \! dx_1
         \int_{0}^{(1-F)\psi} \! dx_3
         \int_{0}^{(1-F)\psi} \! dx_4  \left(
         \frac{1}{[ \mbox{ \ } ]^2 [ \mbox{ \ } ] [ \mbox{ \ } ] }
       + \frac{2}{[ \mbox{ \ }] [ \mbox{ \ }]^2 [ \mbox{ \ }]}
       + \frac{1}{[ \mbox{ \ }] [ \mbox{ \ }] [ \mbox{ \ }]^2} \right) \, ,
\label{eq:3deriv}
\end{equation}
where the bracket contents correspond---in order of appearance---
to those of Eq.~(\ref{eq:2deriv}).  $I_{\rho \rho}$ lacks
the convenient symmetry between first and third terms
that was so handy before.  Nevertheless, integration of the
first term by parts with respect to $x_3$ still helps.
The triple integrals that result cancel the third term
and half the middle term, leaving
\begin{eqnarray}
I_{\rho \rho}  & = &
         \ 2\sum_{p=0}^{1} \sum_{q=0}^{1}
              \frac{(-1)^{p+q}}{A_{p}+B_{q}+1-\rho}
         \int_{0}^{F \psi} \! dx_1
         \int_{0}^{(1-F)\psi} \! dx_4 \,
                \nonumber \\
       &  & \mbox{\hspace{0.6in}} \times
         \frac{1}{[x_4+x_1 +A_{p}+B_{q}+2(2-\rho)]
                      [x_1+B_{q}+1-\rho]} \nonumber \\
       &  & \mbox{} + 2\sum_{p=0}^{1} (-1)^{p}
         \int_{0}^{F \psi} \! dx_1
         \int_{0}^{(1-F)\psi} \! dx_3
         \int_{0}^{(1-F)\psi} \! dx_4 \,
         \frac{1}{[x_3+A_{p}+1-\rho]}  \nonumber \\
       &  & \mbox{\hspace{0.6in}} \times
         \frac{1}{[x_4 + x_3 + x_1 + A_{p} + 2(2-\rho)]^2
                         [x_3 + x_1 + 1-\rho]} \, ,
\label{eq:4deriv}
\end{eqnarray}
where $B_{0} = 0$ and $B_{1} = (1-F)\psi$.

We now
straightforwardly integrate over $x_4$, using
the relation
\begin{displaymath}
\frac{1}{(x+a)(x+b)} = \frac{1}{b-a}
              \left(\frac{1}{x+a} - \frac{1}{x+b} \right).
\end{displaymath}
The result is that
\begin{eqnarray}
I_{\rho \rho}  & = &
         \ 2\sum_{p=0}^{1} \sum_{q=0}^{1} \sum_{r=0}^{1}
              \frac{(-1)^{p+q+r+1}}{A_{p}+B_{q}+1-\rho}
         \int_{0}^{F \psi} \! dx \,
         \frac{\ln [x +A_{p}+B_{q}+B_{r}+2(2-\rho)]}
               {x+B_{q}+1-\rho} \nonumber \\
       &  & \mbox{} + 2\sum_{p=0}^{1} \sum_{q=0}^{1}
         \frac{(-1)^{p+q}}{A_p+B_q+3-\rho}
         \int_{0}^{F \psi} \! dx_1
         \int_{0}^{(1-F)\psi} \! dx_3 \,
         \frac{1}{[x_3+A_{p}+1-\rho]}  \nonumber \\
       &  & \mbox{ \hspace{0.5in} } \times
         \left( \frac{1}{[x_3+x_1+1-\rho]}
               -\frac{1}{[x_3+x_1+A_{p}+B_{q}+2(2-\rho)]} \right).
\label{eq:5deriv}
\end{eqnarray}

Recalling Eq.~(\ref{eq:logint}), we see that, as
$\psi \rightarrow \infty$, the leading part of the first
term in Eq.~(\ref{eq:5deriv}) goes as
$[\ln^2 \psi/(A_{p}+B_{q} + 1-\rho)]$ and
therefore goes to zero unless
$p=q=0$.  The same is true for the second term---which
upon integration over $x_1$ will have a form like that
of the first term.  Hence, we can eliminate the sums
over $p$ and $q$ and, after integrating the second
term over $x_1$, have
\begin{eqnarray}
I_{\rho \rho}  & = &
         \ \frac{2}{1-\rho} \sum_{r=0}^{1}  (-1)^{r+1}
            \int_{0}^{F \psi} \! dx \,
         \frac{\ln [x + B_{r} + 2(2-\rho)]}
               {x + 1-\rho} \nonumber \\
       &  & \mbox{} + \frac{2}{3-\rho} \sum_{r=0}^{1} (-1)^{r+1}
         \int_{0}^{(1-F)\psi} \! dx \,
    \frac{\ln \! \left[ \frac{x+A_{r}+1-\rho}{x+A_{r}+2(2-\rho)} \right]}
              {x + 1-\rho} \, .
\label{eq:6deriv}
\end{eqnarray}

We recognize that the second term is nontrivial only for $r=0$
and apply Eq.~(\ref{eq:logint}) to do the remaining
integrations over $x$.  After dropping terms that go to zero as
$\psi \rightarrow \infty$, we arrive at the ``final'' formula
for $I_{\rho \rho}$:
\begin{eqnarray}
I_{\rho \rho}  & = & \ I_{\rho \rho}^{(1)}
                     + I_{\rho \rho}^{(2)} \, ,
                                 \nonumber \\
       &  & \mbox{\hspace{1in}} \nonumber \\
\left( \frac{1-\rho}{2} \right) I_{\rho \rho}^{(1)} & = & \
        \ln([1-F]\psi)\ln(F\psi) - \frac{1}{2}\ln^2 (F\psi)
          + \sum_{n=1}^{\infty} \frac{(-1)^{n+1}}{n^2}
                       \left( \frac{F}{1-F} \right)^{n}
          - \frac{\pi^2}{6}
                 \nonumber \\
        &   & \mbox{}
          -\ln([1-F]\psi)\ln(1-\rho)
          -\frac{1}{2}\ln^2 (3-\rho) +\ln(3-\rho)\ln(1-\rho)
                 \nonumber \\
        &   & \mbox{}
          + \sum_{n=1}^{\infty} \frac{(-1)^{n+1}}{n^2}
                         \left(\frac{1-\rho}{3-\rho} \right)^{n}
                 \nonumber \\
        &   & \mbox{\hspace{1in} \text{for} \ } F \leq (1-F) ,
                 \nonumber \\
    &  & \mbox{ \hspace{1in} } \nonumber \\
        & = &
        \frac{1}{2}\ln^2 ([1-F]\psi)
          -\sum_{n=1}^{\infty} \frac{(-1)^{n+1}}{n^2}
                   \left(\frac{1-F}{F} \right)^{n}
          -\ln ([1-F]\psi)\ln(1-\rho)
                 \nonumber \\
        &   & \mbox{}
          -\frac{1}{2}\ln^2 (3-\rho) +\ln(3-\rho)\ln(1-\rho)
          + \sum_{n=1}^{\infty} \frac{(-1)^{n+1}}{n^2}
                         \left(\frac{1-\rho}{3-\rho} \right)^{n}
                 \nonumber \\
        &  & \mbox{\hspace{1in} \text{for} \ } F > (1-F) ; \nonumber \\
        &  & \mbox{\hspace{1in}} \nonumber \\
\left( \frac{3-\rho}{2} \right) I_{\rho \rho}^{(2)} & = &
        \frac{\pi^2}{6}
        + \frac{1}{2} \ln^2 \left(\frac{1-\rho}{3-\rho} \right)
          - \sum_{n=1}^{\infty} \frac{(-1)^{n+1}}{n^2}
                         \left(\frac{1-\rho}{3-\rho} \right)^{n} \, .
\label{eq:8derivB}
\end{eqnarray}

Before undoing the differentiations with respect to $\rho$,
we pause to remark on the meaning that can be attached
to the derivatives $I_{\rho \rho}$ and $I_{\rho}$.
The second derivative $I_{\rho \rho}$
can be interpreted physically [after multiplication by
$N_{\text{ch}}U_{2}(t/\delta)^4$] as reflecting
a change in the effective differential charging energy
$U_{\text{eff}} =
     2 [\partial^2 E_{g}^{(0)}(\rho)/\partial \rho^2]_{\rho=0}$,
where $E_{g}^{(0)}(\rho)$ is the ground-state energy as
a function of $\rho$ for a given value of the dimensionless
channel conductance $g$.  (One might choose to
speak of an effective differential capacitance~\cite{Golubev}
$C_{\text{eff}} = e^{2}/2 U_{\text{eff}}$.)
Similarly,~\cite{Golubev,Grabert,Matveev1} up to a
proportionality factor, the first derivative $I_{\rho}$ can
be understood as a tunneling-induced correction
to the expectation value of $\hat{n}$.

What is desired here, however, is $I$ itself, $I$ being
proportional to Diagram 2's contribution to the fourth-order
energy shift [recall Eq.~(\ref{eq:exshft3})].
Integrating $I_{\rho \rho}$ twice with respect to $\rho$
gives us $I$ up to additive terms that are constant or
linear with respect to $\rho$:
\begin{equation}
I(\rho,F,\psi) = a_0 + a_1 \rho +
          \int_{0}^{\rho} \! dx_1 \! \int_{0}^{x_1} \! dx_2 \,
                   I_{\rho \rho} (x_2,F,\psi) \, .
\label{eq:cong1}
\end{equation}
As mentioned in Sec. II and at the beginning of this
appendix, the unknown terms $(a_0 + a_1 \rho)$ are
not relevant to our result.  The $a_1 \rho$ term is
negligible due to the existence of the mirror image of
Diagram 2, in which the roles of dots 1 and 2 are
exchanged.  Such a
switch of $\hat{n}_1$ and $\hat{n}_2$ is equivalent
---in calculating energies---to taking
$\rho \rightarrow -\rho$.  Consequently, when the total
fourth-order shift is calculated,
the $a_1 \rho$ in Eq.~(\ref{eq:cong1}) cancels
with the $-a_1 \rho$ from the mirror image.
Likewise, the $a_0$ part drops from the final
result as we are only concerned with the difference
between the energy shifts for arbitrary $\rho$ and
$\rho = 0$.

The irrelevance of the $a_0$ and $a_1 \rho$ terms
tells us that we need only calculate
$I$ modulo terms constant or linear with respect to $\rho$.
In other words, we need only find an equivalence class
\begin{equation}
I(\rho,F,\psi) \cong \int_{0}^{\rho} \! dx_1 \!
                     \int_{0}^{x_1} \! dx_2 \,
                   I_{\rho \rho} (x_2,F,\psi) \, ,
\label{eq:cong2}
\end{equation}
where the congruence symbol indicates
equivalence up to additive terms that are constant or
linear with respect to $\rho$.  We are therefore
free to drop any constant or linear terms
that crop up on the right side of Eq.~(\ref{eq:cong2}).

Confident that we have figured out what we
wish to do, we can return to the pedestrian
business of doing it.  We observe that the
$\rho$-dependent sum in Eq.~(\ref{eq:8derivB})
can be written in a more integrable form:
\begin{eqnarray}
\sum_{n=1}^{\infty} \frac{(-1)^{n+1}}{n^2}
            \left(\frac{1-\rho}{3-\rho} \right)^{n}
   & = & \ \sum_{n=1}^{\infty} \frac{(-1)^{n+1}}{n^2}
            \left(\frac{1}{3} \right)^{n}
    + \ln 2 [\ln (1-\rho) - \ln (3-\rho) + \ln 3]
     \nonumber \\
   &   & \mbox{}
    + \int_{0}^{\rho} \! dx \,
        \ln \! \left(\frac{2-x}{3-x} \right)
        \left( \frac{-1}{1-x} + \frac{1}{3-x} \right) \, .
\label{eq:sum3}
\end{eqnarray}
Integration of $I_{\rho \rho}^{(1)}$ and $I_{\rho \rho}^{(2)}$
with respect to $\rho$ gives
\begin{eqnarray}
\frac{1}{2} I_{\rho}^{(1)}
       & = & -\ln (1-\rho) \left[
        \ln([1-F]\psi)\ln(F\psi) - \frac{1}{2}\ln^2 (F\psi) \right.
          + \sum_{n=1}^{\infty} \frac{(-1)^{n+1}}{n^2}
                       \left( \frac{F}{1-F} \right)^{n}
     \nonumber \\
       &   & \mbox{\hspace{0.6in}}    \left.
          - \frac{\pi^2}{6}
          + \sum_{n=1}^{\infty} \frac{(-1)^{n+1}}{n^2}
               \left(\frac{1}{3} \right)^{n}
          + \ln 2 \ln 3                    \right]
                  \nonumber \\
        &   & \mbox{}
          +\frac{1}{2} \ln^2 (1-\rho)
                   [ \ln([1-F]\psi) - \ln 2 ]
          -\frac{1}{2}\int_{0}^{\rho} \! dx \,
                  \frac{\ln^2 (3-x)}{1-x}
              \nonumber \\
        &   & \mbox{}
          + \int_{0}^{\rho} \! dx \,
                  \frac{\ln(3-x)\ln(1-x)}{1-x}
          - \ln 2 \int_{0}^{\rho} \! dx \, (\rho-x)
                   \frac{\ln (3-x)}{1-x}
              \nonumber \\
        &   & \mbox{}
          + \int_{0}^{\rho} \! dx_1 \, \frac{\rho-x_1}{1-x_1}
                 \int_{0}^{x_1} \! dx_2 \,
                     \ln \! \left(\frac{2-x_2}{3-x_2} \right)
         \left( \frac{-1}{1-x_2} + \frac{1}{3-x_2} \right)
                 \nonumber \\
        &   & \mbox{\hspace{1in} \text{for} \ } F \leq (1-F),
                 \nonumber \\
    &   & \mbox{\hspace{1in}} \nonumber \\
       & = & -\ln (1-\rho) \left[
            \frac{1}{2}\ln^2 ([1-F]\psi)
          - \sum_{n=1}^{\infty} \frac{(-1)^{n+1}}{n^2}
                       \left( \frac{1-F}{F} \right)^{n} \right.
       \nonumber \\
       &   & \mbox{\hspace{0.6in}} \left.
          + \sum_{n=1}^{\infty} \frac{(-1)^{n+1}}{n^2}
               \left(\frac{1}{3} \right)^{n}
          + \ln 2 \ln 3                    \right]
                   + \ldots
                 \nonumber \\
        &  & \mbox{\hspace{1in} \text{for} \ } F > (1-F); \nonumber \\
        &  & \mbox{\hspace{1in}} \nonumber \\
\frac{1}{2} I_{\rho}^{(2)}
               & = & \
        \left[-\ln (3-\rho) + \ln 3 \right] \left[\frac{\pi^2}{6}
          - \sum_{n=1}^{\infty} \frac{(-1)^{n+1}}{n^2}
               \left(\frac{1}{3} \right)^{n}
          - \ln 2 \ln 3                    \right]
              \nonumber \\
        &   & \mbox{}
          - \frac{1}{6} [ \ln^3 (3-\rho) - \ln^3 3 ]
          - \frac{1}{2}\ln 2 [\ln^2 (3-\rho) - \ln^2 3 ]
              \nonumber \\
        &   & \mbox{}
          + \frac{1}{2}\int_{0}^{\rho} \! dx \,
                \frac{\ln^2 (1-x)}{3-x}
          - \int_{0}^{\rho} \! dx \,
                \frac{\ln (3-x) \ln (1-x)}{3-x}
              \nonumber \\
        &   & \mbox{}
          -\ln 2 \int_{0}^{\rho} \! dx \, \frac{\ln (1-x)}{3-x}
              \nonumber \\
        &   & \mbox{}
          - \int_{0}^{\rho} \! dx_1 \, \frac{1}{3-x_1}
               \int_{0}^{x_1} \! dx_2 \,
                    \ln \! \left(\frac{2-x_2}{3-x_2} \right)
               \left( \frac{-1}{1-x_2} + \frac{1}{3-x_2} \right) .
\label{eq:1deriv1}
\end{eqnarray}
The ellipsis in the second equation for
$I_{\rho}^{(1)}$ indicates that the remainder of
$I_{\rho}^{(1)}$ for the system above half-filling is
the same as the corresponding remainder for the system
below or at half-filling.

In deriving Eq.~(\ref{eq:1deriv1}), we eliminated a number of
integrals over $x_i$'s by using an identity~\cite{Press}
that is easily derived for double integrals:
\begin{equation}
\int_{0}^{\rho} \! dx_1 \! \int_{0}^{x_1} \! dx_2 \, f(x_2)
   = \int_{0}^{\rho} \! dx \, (\rho-x) f(x) \, .
\end{equation}
Nonetheless, in the final terms of $I_{\rho}^{(1)}$ and
$I_{\rho}^{(2)}$, double integrals remain.  These can
be reduced to single-integral form with a little extra
work.  Defining $L^{(m)}$ to be the last term of
$(1/2)I_{\rho}^{(m)}$, we discover that
\begin{eqnarray}
L^{(1)}
        & = & \int_{0}^{\rho} \! dx_2  \,
                   \ln \! \left(\frac{2-x_2}{3-x_2} \right)
                   \left(\frac{-1}{1-x_2} + \frac{1}{3-x_2} \right)
                \int_{x_2}^{\rho} \! dx_1 \,
                   \left( 1 - \frac{1-\rho}{1-x_1} \right)
        \nonumber \\
        &   & \mbox{\hspace{1in}} \nonumber \\
        & = & \int_{0}^{\rho} \! dx \,
                   \ln \! \left(\frac{2-x}{3-x} \right)
                   \left(\frac{-1}{1-x} + \frac{1}{3-x} \right)
        \nonumber \\
        &   & \mbox{ \hspace{0.6in} } \times
 \left[ (\rho-x) + (1-\rho)\ln \! \left(\frac{1-\rho}{1-x} \right) \right] ,
                            \nonumber \\
        &   & \mbox{\hspace{1in}} \nonumber \\
L^{(2)} & = &  -\int_{0}^{\rho} \! dx \,
                   \ln \! \left(\frac{2-x}{3-x} \right)
                   \left(\frac{-1}{1-x} + \frac{1}{3-x} \right)
        \nonumber \\
        &   & \mbox{ \hspace{0.6in} } \times
    \left[ (\rho-x) + (3-\rho)\ln \! \left(\frac{3-\rho}{3-x} \right) \right].
\label{eq:L1L2}
\end{eqnarray}

Now we perform the final integration over $\rho$,
deriving $I^{(1)}$ and $I^{(2)}$,
where $I \cong I^{(1)} + I^{(2)}$.
\begin{eqnarray}
\frac{1}{2} I^{(1)}
       & \cong & \ (1-\rho)\ln (1-\rho) \left[
        \ln([1-F]\psi)\ln(F\psi) - \frac{1}{2}\ln^2 (F\psi)
          + \sum_{n=1}^{\infty} \frac{(-1)^{n+1}}{n^2}
                       \left( \frac{F}{1-F} \right)^{n}  \right.
     \nonumber \\
       &   & \mbox{\hspace{0.6in}} \left.
          - \frac{\pi^2}{6}
          + \sum_{n=1}^{\infty} \frac{(-1)^{n+1}}{n^2}
               \left(\frac{1}{3} \right)^{n}
          + \ln 2 \ln 3
          + \ln([1-F]\psi) - \ln 2 \right]
        \nonumber \\
        &   & \mbox{}
          -\frac{1}{2} (1-\rho)\ln^2 (1-\rho)
                   [ \ln([1-F]\psi) - \ln 2 ]
          -\frac{1}{2}\int_{0}^{\rho} \! dx \, (\rho-x)
                  \frac{\ln^2 (3-x)}{1-x}
              \nonumber \\
        &   & \mbox{}
          + \int_{0}^{\rho} \! dx \, (\rho-x)
                  \frac{\ln(3-x)\ln(1-x)}{1-x}
          - \ln 2 \int_{0}^{\rho} \! dx \, (\rho-x)
                   \frac{\ln (3-x)}{1-x}
              \nonumber \\
        &   & \mbox{}
          + (1-\rho)\int_{0}^{\rho} \! dx \,
                     \ln \! \left(\frac{1-\rho}{1-x} \right)
                     \ln \! \left(\frac{2-x}{3-x} \right)
            \left( \frac{-1}{1-x} + \frac{1}{3-x} \right)
                 \nonumber \\
        &   & \mbox{}
          + \int_{0}^{\rho} \! dx \, (\rho-x)
                     \ln \! \left(\frac{2-x}{3-x} \right)
         \left( \frac{-1}{1-x} + \frac{1}{3-x} \right)
                 \nonumber \\
        &   & \mbox{\hspace{1in} \text{for} \ } F \leq (1-F),
                 \nonumber \\
    &  & \mbox{ \hspace{1in} } \nonumber \\
    & \cong & \ (1-\rho)\ln (1-\rho) \left[
            \frac{1}{2}\ln^2 ([1-F]\psi)
          - \sum_{n=1}^{\infty} \frac{(-1)^{n+1}}{n^2}
                       \left( \frac{1-F}{F} \right)^{n}  \right.
       \nonumber \\
       &   & \mbox{\hspace{0.6in}} \left.
          + \sum_{n=1}^{\infty} \frac{(-1)^{n+1}}{n^2}
               \left(\frac{1}{3} \right)^{n}
          + \ln 2 \ln 3
          + \ln([1-F]\psi) - \ln 2      \right] + \ldots
                  \nonumber \\
        &  & \mbox{\hspace{1in} \text{for} \ } F > (1-F);  \nonumber \\
   & & \mbox{\hspace{1in}} \nonumber \\
\frac{1}{2} I^{(2)}
               & \cong & \
        [(3-\rho)\ln (3-\rho) - 3\ln 3 ] \left[\frac{\pi^2}{6}
          - \sum_{n=1}^{\infty} \frac{(-1)^{n+1}}{n^2}
               \left(\frac{1}{3} \right)^{n}
          - \ln 2 \ln 3                    \right]
              \nonumber \\
        &   & \mbox{}
          + \frac{1}{6} \left[ (3-\rho)\ln^3 (3-\rho)
                         -3(3-\rho)\ln^2 (3-\rho)
                         +6(3-\rho)\ln (3-\rho) \right.
              \nonumber \\
        &   & \mbox{\hspace{0.6in}} \left.
                         -3\ln^3 3 + 9\ln^2 3 -18\ln 3 \right]
                 \nonumber \\
        &   & \mbox{}
          + \frac{1}{2}\ln 2 [(3-\rho)\ln^2 (3-\rho)
                              -2(3-\rho)\ln (3-\rho)
                              -3\ln^2 3 + 6\ln 3 ]
              \nonumber \\
        &   & \mbox{}
          + \frac{1}{2}\int_{0}^{\rho} \! dx \, (\rho-x)
                \frac{\ln^2 (1-x)}{3-x}
          - \int_{0}^{\rho} \! dx \, (\rho-x)
                \frac{\ln (3-x) \ln (1-x)}{3-x}
              \nonumber \\
        &   & \mbox{}
          -\ln 2 \int_{0}^{\rho} \! dx \, (\rho-x)
                      \frac{\ln (1-x)}{3-x}
              \nonumber \\
        &   & \mbox{}
          -(3-\rho) \int_{0}^{\rho} \! dx \,
                    \ln \! \left(\frac{3-\rho}{3-x} \right)
                    \ln \! \left(\frac{2-x}{3-x} \right)
               \left( \frac{-1}{1-x} + \frac{1}{3-x} \right)
               \nonumber \\
        &   & \mbox{}
          - \int_{0}^{\rho} \! dx \, (\rho-x)
                    \ln \! \left(\frac{2-x}{3-x} \right)
               \left( \frac{-1}{1-x} + \frac{1}{3-x} \right).
\label{eq:finish1}
\end{eqnarray}

We are essentially done.  Upon multiplying the sum of
$I^{(1)}$ and $I^{(2)}$ by
$N_{\text{ch}}U_{2}(t/\delta)^4$, we have the relevant
contribution from Diagram 2 to the fourth-order energy
shift.  After so much work, one might wonder whether
we have achieved anything more.  Providentially, the
answer is that, yes, we have.
As explained earlier, we have also solved
for the contribution from the corresponding mirror-image
diagram, which is obtained by replacing $\rho$ with
$-\rho$ in Eq.~(\ref{eq:finish1}).  Perhaps more surprisingly,
we have solved for the contributions from another
pair of mirror-image terms.  A swap of $F$ and $(1-F)$ in
Eq.~(\ref{eq:exshft3}) turns it into the formula for
the contribution from Diagram~3 of Fig.~2(b).  Thus,
exchanging $F$ and $(1-F)$ in Eq.~(\ref{eq:finish1})
yields the contribution from Diagram~3.  A further
replacement of $\rho$ with $-\rho$ gives the contribution
from Diagram~3's mirror image.
The cost of calculating Diagram~2 is high, but at least
we benefit from a package deal---4 for the price of 1.

\section{Details of the Strong-Coupling Calculation}

This appendix consists of three parts presenting
various calculations described or cited in Sec. III.
The first part computes $S_{\text{b}}^{(1)}$,
$S_{\text{b}}^{(2)}$, $K_{\text{c}}(\tau)$, and
$K_{\text{c}}(0)$, thereby producing
the results quoted in Eqs.~(\ref{eq:moments})
and making explicit the origin of the
factor $e^{\gamma}$ that appears in the prefactors
of Eqs.~(\ref{eq:Hnewferm}) and (\ref{eq:Del2str}).
The second part of the appendix provides the derivation of
the first strong-coupling energy correction
[see Eq.~(\ref{eq:Del1str})].
The third part derives the second strong-coupling correction
[see Eq.~(\ref{eq:Del2str})].

\subsection{Calculation of Charge-Channel Averages}

The leap from Eq.~(\ref{eq:effact3})
to Eq.~(\ref{eq:moments}) in
Sec.~III requires evaluation of the expectation values
\begin{eqnarray}
D_{1} & = & \left\langle
        \cos \left[\sqrt{\pi} \theta_{\text{c}}(\tau)
                        +\frac{\pi \rho}{2} \right]
            \right\rangle_{\text{c}} \, ,
                             \nonumber \\
D_{2} & = &  \left\langle
          \cos \left[\sqrt{\pi} \theta_{\text{c}}(\tau_1)
                      +\frac{\pi \rho}{2} \right]
          \cos \left[\sqrt{\pi} \theta_{\text{c}}(\tau_2)
                      +\frac{\pi \rho}{2} \right]
             \right\rangle_{\text{c}} \, .
\label{eq:defD}
\end{eqnarray}
[Recall that time-ordering is implicit in the
path-integral definition of
$\langle \hat{A} \rangle_{\text{c}}$ in
Eq.~(\ref{eq:chargeave}).]
The cosines and products of cosines can be
written as linear combinations of terms of the
form $e^{\hat{Z}}$, where $\hat{Z}$ is linear in the
charge displacement operators $\theta_{\text{c}}(\tau)$
and the charge displacement operators are
themselves linear in boson creation and
annihilation operators (see Ref.~21).
Therefore, one can apply a standard relation for the
expectation value of the exponential of a linear combination
of boson operators~\cite{Emery}
\begin{equation}
\langle e^{\hat{Z}} \rangle
   = e^{\frac{1}{2} \langle \hat{Z}^2 \rangle} \, ,
\label{eq:expbose}
\end{equation}
which can easily be shown to hold for our charge-integration
brackets with implicit time-ordering.

Using Eq.~(\ref{eq:expbose}), we discover that
\begin{eqnarray}
D_{1} & = & e^{-\frac{\pi}{2} K_{\text{c}}(0)}
            \cos \! \left(\frac{\pi \rho}{2}\right) \, ,
                    \nonumber \\
D_{2} & = &
  \frac{1}{2} e^{-\pi [K_{\text{c}}(0) + K_{\text{c}}(\tau_1-\tau_2)]}
               \cos (\pi \rho)
 +\frac{1}{2} e^{-\pi [K_{\text{c}}(0) - K_{\text{c}}(\tau_1-\tau_2)]} \, ,
\label{eq:D1D2}
\end{eqnarray}
where $K_{\text{c}}(\tau)$ is the charge-charge correlation function,
\begin{equation}
K_{\text{c}}(\tau) = \left\langle \theta_{\text{c}}(\tau)
                       \theta_{\text{c}}(0) \right\rangle_{\text{c}} \, .
\label{eq:Kc1}
\end{equation}
Eq.~(\ref{eq:moments}) of Sec. III follows immediately.
We find $S_{\text{b}}^{(1)}$ by replacing
$\cos[\sqrt{\pi} \theta_{\text{c}}(\tau) + \pi \rho/2]$
in $S_{\text{b}}$ with $D_{1}$.  For $S_{\text{b}}^{(2)}$,
we recall that
\begin{displaymath}
\langle [S_{\text{b}}
  -\langle S_{\text{b}} \rangle_{\text{c}} ]^2 \rangle_{\text{c}}
= \langle S_{\text{b}}^2 \rangle_{\text{c}}
    - \langle S_{\text{b}} \rangle_{\text{c}}^2
\end{displaymath}
and apply the formulas for $D_{1}$ and $D_{2}$
accordingly.

To get the formula for $K_{\text{c}}(\tau)$
[Eq.~(\ref{eq:Kcorrel})], we must labor a bit more.
Because the unperturbed action $S_{0}^{\text{(c)}}$ is
quadratic in charge displacement operators
$\tilde{\theta}_{\text{c}}(\omega_{m})$,
$S_{0}^{\text{(c)}}$ fits exactly the form for the
canonical action of a real scalar field.~\cite{Amit}
Consequently,
\begin{equation}
\left\langle \tilde{\theta}_{\text{c}}(\omega_{m})
               \tilde{\theta}_{\text{c}}(-\omega_{n})
                       \right\rangle_{\text{c}}
    = \frac{\beta}{|\omega_{m}| + \frac{2 U_{2}}{\pi} }
             \, \delta_{\omega_{m}, \omega_{n}} \, .
\label{eq:thetasqr}
\end{equation}
{}From this identity and the relation between $\theta_{\text{c}}(\tau)$
and its Fourier transform [recall Eq.~(\ref{eq:thetaft})],
we construct a summation formula for $K_{\text{c}}(\tau)$:
\begin{equation}
K_{\text{c}}(\tau) = \frac{1}{\beta} \sum_{\omega_{m}}
                       \frac{e^{-i \omega_{m} \tau}}
                  { |\omega_{m}| + \frac{2 U_{2}}{\pi} }.
\label{eq:Kc2}
\end{equation}

In the zero-temperature ($\beta \rightarrow \infty$) limit,
we may safely transform this sum into an integral.
Before doing so, however, we should note that, unless the
$\omega_{m}$ possess an ultraviolet cut-off, $K_{\text{c}}(0)$
diverges logarithmically.
The standard means of imposing such a cut-off in Luttinger
liquid theory~\cite{Emery} is to insert a factor of
$e^{-|\omega_{m}|/W}$ on the right side of
Eq.~(\ref{eq:thetaft}).
This insertion generates a factor of
$e^{-2 |\omega_{m}|/W}$ in Eq.~(\ref{eq:Kc2}), yielding
\begin{equation}
K_{\text{c}}(\tau) = \int_{-\infty}^{\infty} \! \frac{d\omega}{2\pi} \,
                       \frac{e^{-i \omega \tau}
                              e^{-\frac{2 |\omega|}{W}} }
                  { |\omega| + \frac{2 U_{2}}{\pi} } \, ,
\label{eq:Kc3}
\end{equation}
which is equivalent to Eq.~(\ref{eq:Kcorrel}) in Sec.~III.

The way is clear for evaluation of the same-time corrrelation
function $K_{\text{c}}(0)$.
After setting $\tau = 0$ in Eq.~(\ref{eq:Kc3}), we
integrate by parts and convert to the dimensionless integration
variable $x = 2 \omega /W$.  The result is that
\begin{eqnarray}
K_{\text{c}}(0) & = &
               -\frac{1}{\pi} \ln \! \left( \frac{4 U_{2}}{\pi W} \right)
               +\frac{1}{\pi} \int_{0}^{\infty} \! dx \,
                    e^{-x} \ln \! \left(x + \frac{4 U_{2}}{\pi W} \right)
                                           \nonumber \\
         & = & -\frac{1}{\pi} \ln \! \left( \frac{4 U_{2}}{\pi W} \right)
               +\frac{1}{\pi} e^{(4 U_{2}/\pi W)}
                   \left( \int_{0}^{\infty} \! dx \, e^{-x} \ln x
                         -\int_{0}^{4 U_{2}/\pi W} \! dx \,
                                    e^{-x} \ln x
                   \right) \, .
\label{eq:Kczero1}
\end{eqnarray}
The first integral in the parentheses equals the negative of
$\gamma$, the Euler-Mascheroni constant.~\cite{Spiegel}  The second
integral goes to zero as we take the limit
$W/U_{2} \rightarrow \infty$.  In this limit, the exponential factor
multiplying the integrals goes to $1$.
The final result is the following:
\begin{equation}
K_{\text{c}}(0) = -\frac{1}{\pi}
           \ln \! \left( \frac{4 e^{\gamma} U_{2}}{\pi W} \right).
\label{eq:Kczero2}
\end{equation}

The derivation of Eq.~(\ref{eq:Kczero2}) shows that
the coefficient $e^{\gamma}$ comes from
exponentiating a secondary part of
$\langle \theta_{\text{c}}(0) \theta_{\text{c}}(0) \rangle$.
One might be concerned that
Luttinger liquid theory does not faithfully capture
such subsidiary dependences.~\cite{Solyom}
However, Sec. IV presents evidence that these coefficients
are general and independent of the high-energy band structure.

\subsection{The First Strong-Coupling Correction}

As stated in Sec. III, in the limit of strong coupling
($g \rightarrow 1$), the first correction
[see Eq.~(\ref{eq:Del1str})] to the
open-channel ($g = 0$) ground-state energy is obtained
by diagonalizing the Hamiltonian $H_{\text{New}}$
[see Eq.~(\ref{eq:Hnew})].  This diagonalization can
be accomplished through another version of the
``debosonization'' procedure used by Matveev.~\cite{Matveev2}
As we wish to ``debosonize'' the action
$S_{\text{New}} = S_{0}^{\text{(s)}} + S_{\text{b}}^{(1)}$
[recall Eqs.~(\ref{eq:strongmod2}) and (\ref{eq:moments})],
it is useful to
observe that $S_{0}^{\text{(s)}}$ corresponds to the
Euclidean action for non-interacting fermions on a
semi-infinite lattice ending at $x=0$.~\cite{Kane}
For these fermions, we take $\theta_{\text{s}}(\tau)$
to correspond to the $x=0$ value of the phase field,
$\phi_f (\tau) = \Phi_f (0,\tau)$, rather than
the $x=0$ value of the charge displacement field
$\theta_f(\tau) = \Theta_f (0,\tau)$.
Making $\theta_f (\tau) = 0$ the boundary condition at
the edge, we find that the
properly normalized creation operator for a fermion at
$x=0$ is given by
\begin{equation}
\psi_f^{\dagger} (0,\tau) = \sqrt{\frac{W}{4 \pi \hbar v_F}}
     \, e^{i \sqrt{\pi} \phi_f (\tau)} \, ,
\label{eq:fermibose}
\end{equation}
where, as usual, $W$ is the bandwidth and $v_F$ is the
Fermi velocity.~\cite{Kane,Emery}
$\psi_f^{\dagger}(0,\tau)$ can be
expressed in terms of reciprocal-space creation operators:
\begin{equation}
\psi_f^{\dagger}(0,\tau) = \frac{1}{\sqrt{2 \pi}}
   \int_{-\Lambda}^{\Lambda} \! dk \, f^{\dagger}_{k} \, .
\label{eq:fermifermi}
\end{equation}
The fermionic energies are cut off in the usual way at
$W/2$, the corresponding wave-vector cut-off being
$\Lambda = W/2 \hbar v_F$.

After these machinations, ``refermionization''
proceeds apace.  Since the unperturbed
action $S_{0}^{\text{(s)}}$ is an action for non-interacting
fermions, the unperturbed Hamiltonian $H_{0}^{\text{(s)}}$
is simply the sum of the single-particle energies of those
fermions.  On the other hand, the perturbation
$H_{\text{b}}^{(1)}$ that corresponds to $S_{\text{b}}^{(1)}$
is a term linear in fermion creation and annihilation operators.
In particular, using Eq.~(\ref{eq:Kczero2}) to determine
$e^{-\frac{\pi}{2} K_{\text{c}}(0)}$, we obtain
\begin{eqnarray}
H_{0}^{\text{(s)}} & = & \int_{-\Lambda}^{\Lambda} \! dk \,
    \xi_k \, f^{\dagger}_{k} f_{k} \, ,
                    \nonumber \\
H_{\text{b}}^{(1)} & = &
     \tilde{V} \cos \! \left(\frac{\pi \rho}{2} \right)
     \sqrt{\frac{2 e^{\gamma} \hbar v_F U_{2}}{\pi^3} }
     \int_{-\Lambda}^{\Lambda} \! dk \, (f^{\dagger}_{k} + f_{k}) \, ,
\label{eq:Hnewferm1}
\end{eqnarray}

Not being quadratic in fermion creation and annihilation
operators, the fermionic Hamiltonian we have derived is not
yet in an easily diagonalizable form.  To make it
so, we follow Matveev~\cite{Matveev2} in defining a
new set of fermion operators such that
\begin{equation}
f_{k} = (d + d^{\dagger})d_{k} \, .
\label{eq:Matvtrans}
\end{equation}
Plugging this equivalence into Eq.~(\ref{eq:Hnewferm1})
yields Eq.~(\ref{eq:Hnewferm}) of Sec.~III.

One can now perform the Bogoliubov transformation
that produces Eq.~(\ref{eq:Hnewdiag}).
To find the correction to the open-channel energy,
one notes that $H_{\text{b}}^{(1)}$ of
Eq.~(\ref{eq:Hnewferm}) has an expectation value of
zero in the ground state of  $H_{0}^{\text{(s)}}$,
which is the open-channel ($\tilde{V}=0$)
part of $H_{\text{New}}$.
Therefore, if the ground state of $H_{0}^{\text{(s)}}$
is represented by the ket $|0\rangle$,
$\langle 0 | H_{\text{New}} | 0 \rangle = E_{0}$,
where $E_{0}$ is the ground-state energy for
$H_{0}^{\text{(s)}}$.
{}From the diagonalized form
of $H_{\text{New}}$ [see Eq.~(\ref{eq:Hnewdiag})],
it is then deduced that the equation for
$(E_{\text{New}}-E_{0})$
is the following:
\begin{equation}
\Delta_{\text{str}}^{(1)}(\rho) = -\int_{0}^{\Lambda} \! dk \, \xi_k
          \langle 0 | C_{k}^{\dagger} C_{k}
                     +\tilde{C}_{k}^{\dagger} \tilde{C}_{k}
                                | 0 \rangle \, .
\label{eq:shft1A}
\end{equation}

At this point, it is necessary to know the exact equations for
$\tilde{C}_{k}$ and $C_{k}$.
As found by Matveev,~\cite{Matveev2}
for $\Gamma = \tilde{V}^2  [8 e^{\gamma} U_{2}/\pi^2]
               \cos^2 (\pi \rho/2)$, they are
\begin{eqnarray}
\tilde{C}_{k} & = & \frac{d_{k} + d_{-k}^{\dagger}}{\sqrt{2}} \, ,
                               \nonumber \\
C_{k}         & = & \frac{\xi_{k}}{\sqrt{\xi_{k}^{2} + \Gamma^2}}
                       \frac{d_{k} - d_{-k}^{\dagger}}{\sqrt{2}}
     -\sqrt{\frac{\hbar v_F \Gamma}{2 \pi (\xi_{k}^{2} + \Gamma^2)} }
                    (d + d^{\dagger})
                               \nonumber \\
              &   & \mbox{\hspace{0.3in}}
                   +\frac{\Gamma}{\pi \sqrt{\xi_{k}^{2} + \Gamma^2}}
                      \,  {\cal P} \! \! \int_{-\Lambda}^{\Lambda}
                    \frac{d\xi_{k^{\prime}}}{\xi_{k} - \xi_{k^{\prime}}}
      \,  \frac{d_{k^{\prime}} - d_{-k^{\prime}}^{\dagger}}{\sqrt{2}} \, ,
\label{eq:Bogfull}
\end{eqnarray}
As before, the symbol $\cal P$ indicates that only the
principal value of the integral is computed.

With the explicit equations for $\tilde{C}_{k}$ and $C_{k}$
before us, it is clear that, for $k > 0$,
$\tilde{C}_{k} |0\rangle = 0$, and
\begin{equation}
 \Delta_{\text{str}}^{(1)}(\rho)  =
            -\int_{0}^{\Lambda} \! dk \, \xi_k
          \langle 0 | C_{k}^{\dagger} C_{k} | 0 \rangle.
\label{eq:shft1B}
\end{equation}
Concentrating on what remains, we see that,
for $k > 0$, both the first term of $C_{k}$ and the
$k^{\prime} > 0$ part of the third term of $C_{k}$
annihilate the $H_{0}^{\text{(s)}}$ ground
state.  Hence,
\begin{eqnarray}
\langle 0| C_{k}^{\dagger} C_{k} |0 \rangle
   & = & \frac{\hbar v_F \Gamma}{2 \pi (\xi_{k}^{2} + \Gamma^2)}
          \left[ 1 + \frac{\Gamma}{\pi \hbar v_F}
            \int_{0}^{\Lambda} \!
                 \frac{dk^{\prime}}{k+k^{\prime}} \!
            \int_{0}^{\Lambda} \!
                 \frac{dk^{\prime \prime}}{k+k^{\prime \prime}} \,
        \langle 0| (d_{-k^{\prime}}^{\dagger} + d_{k^{\prime}})
        (d_{-k^{\prime \prime}} + d_{k^{\prime \prime}}^{\dagger})
                            |0 \rangle
           \right]
                    \nonumber \\
   & = & \frac{\hbar v_F \Gamma}{2 \pi (\xi_{k}^{2} + \Gamma^2)}
        +\frac{\Gamma^2}{\pi^2 (\xi_{k}^{2} + \Gamma^2)}
        \left( \frac{1}{k} - \frac{1}{k + \Lambda} \right) .
\label{eq:numbexp}
\end{eqnarray}

Plugging into Eq.~(\ref{eq:shft1B}), we find that
\begin{eqnarray}
\Delta_{\text{str}}^{(1)}(\rho)  & = &
   -\frac{\Gamma}{2\pi} \int_{0}^{W/2} \!
            \frac{\xi_k \, d\xi_k}{\xi_{k}^{2} + \Gamma^2}
   -\frac{\Gamma^2}{\pi^2} \int_{0}^{W/2} \!
            \frac{d\xi_k}{\xi_{k}^{2} + \Gamma^2}
                                 \nonumber \\
   &  & \mbox{\hspace{0.5in}}
   +\frac{\Gamma^2}{\pi^2} \int_{0}^{W/2} \!
            \frac{\xi_k \, d\xi_k}
                 {(\xi_k + \frac{W}{2})(\xi_{k}^{2} + \Gamma^2)}
                                 \nonumber \\
   & = &
   -\frac{\Gamma}{4\pi} \ln \! \left(\frac{W^2}{4\Gamma^2} + 1 \right)
      - \frac{\Gamma}{2\pi}
                                 \nonumber \\
   & = &
   -\frac{\Gamma}{2\pi} \left[ \ln \! \left(\frac{W}{2\Gamma} \right)
                              + 1 \right] .
\label{eq:shft1C}
\end{eqnarray}
Here we have dropped terms that vanish as
$W/U_{2} \rightarrow \infty$.  Eq.~(\ref{eq:Del1str}) is obtained
by applying the identity
$\Gamma = \tilde{V}^2 [8 e^{\gamma} U_{2}/\pi^2] \cos^2 (\pi \rho/2)$.

\subsection{The Second Strong-Coupling Correction}

The second correction term in the strong-coupling limit
[see Eq.~(\ref{eq:Del2str})] is derived by
treating $S_{\text{b}}^{(2)}$ [see Eq.~(\ref{eq:moments})]
as a perturbation to the
system described by $H_{\text{New}}$ of Eq.~(\ref{eq:Hnew}).
Using the standard formula for the grand-canonical potential
in the finite-temperature path-integral approach,~\cite{Negele}
\begin{equation}
\Omega - \Omega_{0}
  = -\frac{1}{\beta} \sum (\mbox{All connected graphs}) \, ,
\label{eq:Omega}
\end{equation}
we see that the lowest-order correction to the ground state
energy of $H_{\text{New}}$ is given by
\begin{equation}
\Delta_{\text{str}}^{(2)}(\rho)
   = \lim_{\beta \rightarrow \infty} \frac{1}{\beta}
  \langle \text{New}| S_{\text{b}}^{(2)} |\text{New} \rangle \, ,
\label{eq:shft2A}
\end{equation}
where $|\text{New} \rangle$ is the ground-state ket for
$H_{\text{New}}$.
The minus sign in Eq.~(\ref{eq:Omega}) has been canceled
by the minus sign that arises from the fact that this leading
term from $S_{\text{b}}^{(2)}$ corresponds to a first-order
graph and therefore carries a factor of $-1$.~\cite{Negele}

Recalling Eq.~(\ref{eq:moments}) and observing
that the parts of $S_{\text{b}}^{(2)}$ that are independent
of $\rho$ are irrelevant to calculation of the
fractional peak splitting $f$, our immediate task is
to evaluate the quantity
\begin{equation}
X(\tau_1,\tau_2) = \left( \frac{\tilde{V} W}{\pi} \right)^2
                     e^{-\pi K_{\text{c}}(0)}
                     \cos^2 \! \left(\frac{\pi \rho}{2} \right)
\langle \text{New}| \cos [\sqrt{\pi} \theta_{\text{s}}(\tau_1)]
    \cos [\sqrt{\pi} \theta_{\text{s}}(\tau_1)] |\text{New} \rangle \, .
\label{eq:defX}
\end{equation}
Under ``debosonization'' (see Part~2 of this appendix), this becomes
\begin{eqnarray}
X(\tau_1,\tau_2)
 & = & \lambda^2
       \int_{-\Lambda}^{\Lambda} \! dk_1 \!
       \int_{-\Lambda}^{\Lambda} \! dk_2 \,
       \langle \text{New}|
         \left[ d_{k_1}^{\dagger}(d + d^{\dagger})
                   + (d + d^{\dagger}) d_{k_1} \right]_{\tau_1}
                                     \nonumber \\
 &   & \mbox{\hspace{0.3in}} \times
         \left[ d_{k_2}^{\dagger}(d + d^{\dagger})
                   + (d + d^{\dagger}) d_{k_2} \right]_{\tau_2}
                             |\text{New} \rangle \, ,
\label{eq:X1}
\end{eqnarray}
where the bracket subscripts indicate that the
enclosed operators are
evaluated at imaginary times $\tau_1$ and $\tau_2$, respectively,
and we have used
\begin{equation}
\lambda = \tilde{V} \cos(\pi \rho/2)
                  \sqrt{2 e^{\gamma} \hbar v_F U_{2} /\pi^3} \, .
\label{eq:lambda}
\end{equation}

We are now within hailing distance of Eq.~(\ref{eq:Del2str}).
Using the truncated equations for $C_{k}$ and $\tilde{C}_{k}$
[recall Eq.~(\ref{eq:bogops})],
we express the $d_k$'s in terms of these operators.
The sub-leading terms in this transformation are negligible
as, in the end result, they take us beyond second order
in $\tilde{V}$.  Similarly, the time-dependence of the operator
sum $(d+d^{\dagger})$ is sub-leading as $(d+d^{\dagger})$
first appears in the expansion of the diagonalizing operators
at order $\tilde{V}$.  Accordingly, $(d+d^{\dagger})$
commutes with $H_{\text{New}}$ to zeroth order and
can be considered time-independent.
In contrast, from Eq.~(\ref{eq:Hnewdiag}), we know that
$C_{k}(\tau) = C_{k} e^{-\xi_k \tau}$ and
$C_{k}^{\dagger}(\tau) = C_{k}^{\dagger} e^{\xi_k \tau}$.
Application of these insights to Eq.~(\ref{eq:X1}) gives
\begin{eqnarray}
X(\tau_1,\tau_2)
 & = &  2 \lambda^2 \int_{0}^{\Lambda} \! dk_1 \!
       \int_{0}^{\Lambda} \! dk_2 \,
    \langle \text{New}| C_{k_1}(\tau_1) C_{k_2}^{\dagger}(\tau_2)
                                 |\text{New} \rangle
                           \nonumber \\
 & = &  \frac{2 \lambda^2}{\hbar v_F} \int_{0}^{W/2} \! d\xi \,
                    e^{-(\tau_1 - \tau_2)\xi}
                           \nonumber \\
 & = & \frac{2 \lambda^2}{\hbar v_F} \,
          \frac{1 - e^{-(\tau_1-\tau_2)W/2}}{\tau_1-\tau_2} \, .
\label{eq:X2}
\end{eqnarray}

We now return to Eqs.~(\ref{eq:moments}) and (\ref{eq:shft2A}).
Switching to dimensionless variables
$x_i = \tau_i W/2$ and substituting for $\lambda$, we obtain
\begin{eqnarray}
\Delta_{\text{str}}^{(2)}(\rho)
   & = & \tilde{V}^2 \cos^2 \! \left(\frac{\pi \rho}{2} \right)
            \frac{8 e^{\gamma} U_{2}}{\pi^3 \beta W}
       \int_{0}^{\beta W/2} \! dx_1 \! \int_{0}^{x_1} \! dx_2 \,
                                   \nonumber \\
   &   & \mbox{\hspace{1.2in}} \times
            \left( 1 - e^{-\pi K_{\text{c}}[2(x_1-x_2)/W]} \right)
            \frac{1 - e^{-(x_1-x_2)}}{x_1-x_2} \, .
\label{eq:shft2B}
\end{eqnarray}
We eliminate one of the integrations by expressing the
integrand in terms of \mbox{$x = (x_1-x_2)$} and observing that
in the double-integral the density of states for a given
value of $x$ is $(\beta W/2 - x)$:
\begin{eqnarray}
\Delta_{\text{str}}^{(2)}(\rho)
   & = & \tilde{V}^2 \cos^2 \! \left(\frac{\pi \rho}{2} \right)
            \frac{8 e^{\gamma} U_{2}}{\pi^3 \beta W}
       \int_{0}^{\beta W/2} \! dx \, \left(\frac{\beta W}{2} - x \right)
            \left[ 1 - e^{-\pi K_{\text{c}}(2 x /W)} \right]
            \frac{1 - e^{-x}}{x} \, .
\label{eq:shft2C}
\end{eqnarray}

Transformation of Eq.~(\ref{eq:shft2C}) into
Eq.~(\ref{eq:Del2str}) follows recognition of the fact
that, for $x$ on the order
of $\beta W/2$, the integrand is effectively zero.
This is known from the identity
\begin{equation}
K_{\text{c}} \left( \frac{2x}{W} \right) = -\frac{1}{\pi}
        \text{Re} \left\{ e^{(4 U_{2}/\pi W)(1 + ix)}
    \text{Ei} \left[-\left( 4 U_2/\pi W \right) (1 +ix) \right]
              \right\} \, ,
\label{eq:expint1}
\end{equation}
where $\text{Ei}[-z]$ is the first exponential
integral function.~\cite{Grad}  For $z \gg 1$,
$\text{Ei}[-z]$ goes as $e^{-z}/z$.
Therefore, the integrand goes to zero as
$1/x^2$ for $x > \pi W/4 U_{2}$, and the
region $x \gg W/U_{2}$ makes a comparatively negligible
contribution to the integral.
This conclusion corroborates the statement made in Sec. III
that the factor $[1 - e^{-\pi K_{\text{c}}(2x/W)}]$
furnishes an  ultraviolet cut-off on the order of $\psi = W/U_{2}$.
Since we calculate in the limit $\beta \rightarrow \infty$,
we know that $\beta W/2 \gg W/U_{2}$ and, hence,
that the integrand is
effectively zero for $x$ on the order of $\beta W/2$.
We can approximate the weight function
$(\beta W/2 - x)$ by $(\beta W/2)$.  The result is
Eq.~(\ref{eq:Del2str}).

\pagebreak

\begin{multicols}{2}
\narrowtext

\end{multicols}


\begin{references}

\bibitem{Review} For an introduction to ``single-electronics,'' see
  M. A. Kastner, Rev. Mod. Phys. {\bf 64}, 849 (1992); D. V. Averin and
  K. K. Likharev, in {\it Mesoscopic Phenomena in Solids}, edited by
  B. L. Altshuler, P. A. Lee, and R. A. Webb
  (North Holland, Amsterdam, 1991); and several
  articles in {\it Single Charge Tunneling}, Vol. 294 of
  {\it NATO Advanced Study Institute Series B Physics},
  edited by H. Grabert
  and M. H. Devoret (Plenum, New York, 1992).

\bibitem{Waugh1} F. R. Waugh, M. J. Berry, D. J. Mar, R. M. Westervelt,
  K. L. Campman, and A.~C. Gossard, Phys. Rev. Lett.
  {\bf 75}, 705 (1995).

\bibitem{Waugh2} F.~R.~Waugh, M. J. Berry,
  C.~H. Crouch, C. Livermore, D. J. Mar, R. M. Westervelt,
  K. L. Campman, and A. C. Gossard, Phys. Rev. B {\bf 53}, 1413 (1996);
  F. R. Waugh, Ph.D. thesis, Harvard University, 1994.

\bibitem{Golden1} J. M. Golden and B. I. Halperin, Phys. Rev. B
  {\bf 53}, 3893 (1996).

\bibitem{Matveev3} K. A. Matveev, L. I. Glazman, and H. U. Baranger,
Phys. Rev. B {\bf 53}, 1034 (1996).

\bibitem{Matveev4} K. A. Matveev, L. I. Glazman, and H. U. Baranger,
preprint (12/11/95, cond-mat/9512082).

\bibitem{Golubev} D. S. Golubev and A. D. Zaikin, Phys. Rev. B
  {\bf 50}, 8736 (1994).

\bibitem{Grabert} Hermann Grabert, Phys. Rev. B {\bf 50}, 17~364 (1994);
Physica B {\bf 194-196}, 1011-1012 (1994).

\bibitem{Matveev2} K. A. Matveev, Phys. Rev. B {\bf 51}, 1743 (1995).

\bibitem{Molen} L. W. Molenkamp, Karsten Flensberg, and M. Kemerink,
  Phys. Rev. Lett. {\bf 75}, 4282 (1995).

\bibitem{Flensberg} Karsten Flensberg, Physica B {\bf 203}, 432 (1994);
  Phys. Rev. B {\bf 48}, 11~156 (1993).

\bibitem{Zimanyi} G. Falci, J. Heins, Gerd Sch\"{o}n, and
  Gergely T. Zimanyi, Physica B {\bf 203}, 409 (1994);
  G. Falci, Gerd Sch\"{o}n, and Gergely T. Zimanyi, Phys. Rev.
  Lett. {\bf 74}, 3257 (1995).

\bibitem{Panyukov} S. V. Panyukov and A. D. Zaikin, Phys. Rev. Lett.
  {\bf 67}, 3168 (1991); Physics Letters A {\bf 183}, 115 (1993).

\bibitem{Frota} H. O. Frota and Karsten Flensberg, Phys. Rev. B
  {\bf 46}, 15~207 (1992).

\bibitem{Kane} C. L. Kane and M. P. A. Fisher, Phys. Rev. Lett.
  {\bf 68}, 1220 (1992); Phys. Rev. B {\bf 46}, 7268 (1992);
  {\it ibid.} {\bf 46}, 15 233 (1992).

\bibitem{Baym} Gordon Baym, {\it Lectures on Quantum Mechanics}
  (Addison-Wesley, Redwood City, 1973), 113ff.

\bibitem{Fisher} M. P. A. Fisher and Wilhelm Zwerger, Phys. Rev. B
  {\bf 32}, 6190 (1985).

\bibitem{Spiegel} Murray R. Speigel, {\it Mathematical Handbook of
  Formulas and Tables}, Schaum's Outline Series (McGraw-Hill,
  New York, 1992).

\bibitem{Matveev1} L. I. Glazman and K. A. Matveev, Zh. \'{E}ksp. Teor.
  Fiz. {\bf 98}, 1834 (1990) [Sov. Phys. JETP {\bf 71}, 1031 (1990)];
  K. A. Matveev, {\it ibid.} {\bf 99}, 1598 (1991) [{\bf 72}, 892 (1991)]

\bibitem{Press} W. H. Press, S. A. Teukolsky, W. T. Vetterling,
  and B. P. Flannery, {\it Numerical Recipes in C: The Art of
  Scientific Computing}, 2nd ed. (Cambridge University Press,
  Cambridge, 1992), 161.

\bibitem{Haldane} F. D. M. Haldane, Phys. Rev. Lett. {\bf 47},
  1840 (1981); J. Phys. C {\bf 14}, 2585 (1981).

\bibitem{Emery} V. J. Emery, ``Theory of the One-Dimensional
  Electron Gas,'' in {\it Highly Conducting One-Dimensional
  Solids}, edited by J. T. Devreese, R. P. Evrard,
  and V. E. van Doren (Plenum, New York, 1979), 247-303.

\bibitem{Ruzin} I. M. Ruzin, V. Chandrasekhar, E. I. Levin, and
  L. I. Glazman, Phys. Rev. B {\bf 45}, 13~469 (1992); L. I. Glazman
  and V. Chandrasekhar, Europhys. Lett. {\bf 19}, 623 (1992).

\bibitem{Ford} C.J.B. Ford, P.J. Simpson, M. Pepper, D. Kern,
  J.E.F. Frost, D.A. Ritchie, and G.A.C. Jones,
  Nanostructured Materials {\bf 3}, 283 (1993).

\bibitem{Stafford1} C. A. Stafford and S. Das Sarma, Phys. Rev. Lett.
  {\bf 72}, 3590 (1994).

\bibitem{Stafford2} C. A. Stafford and S. Das Sarma, preprint
(5/12/95, cond-mat/9505058).
%\bibitem{Stafford2} C. A. Stafford and S. Das Sarma, unpublished.

\bibitem{Kemerink} M. Kemerink and L.W. Molenkamp,
  Appl. Phys. Lett. {\bf 65}, 1012 (1994).

\bibitem{Tsukada} N. Tsukada, M. Gotoda, M. Nunoshita,
  Phys. Rev. B {\bf 50}, 5764 (1994).

\bibitem{Sakamoto} T. Sakamoto, S.W. Hwang, F. Nihey, Y. Nakamura,
  and K. Nakamura, Jpn. J. Appl. Phys. {\bf 33}, 4876 (1994);
  Superlattices and Microstructures {\bf 16}, 291 (1994).

\bibitem{Klimeck} G. Klimeck, Guanlong Chen, and S. Datta,
  Phys. Rev. B {\bf 50}, 2316 (1994); Guanlong Chen, G. Klimeck,
  S. Datta, Guanha Chen, and W. A. Goddard III, {\it ibid.}
  {\bf 50}, 8035 (1994).

\bibitem{Hofmann} F. Hofmann, T. Heinzel, D.A. Wharam, J.P. Kotthaus,
  G. B\"{o}hm, W. Klein, G. Tr\"{a}nkle, and G. Weimann,
  Phys. Rev. B {\bf 51}, 13 872 (1995).

\bibitem{Haug} R.J. Haug, R.H. Blick, and T. Schmidt, Physica B
  {\bf 212}, 207 (1995).

\bibitem{Vaart} N.C. van der Vaart, S.F. Godijn, Y.V. Nazarov,
  C.J.P.M. Harmans, J.E. Mooij, L.W. Molenkamp, C.T. Foxon,
  Phys. Rev. Lett. {\bf 74}, 4702 (1995).

\bibitem{Crouch} C. H. Crouch, C. Livermore, F. R. Waugh,
  R.~M.~Westervelt, K.~L.~Campman, and
  A.~C. Gossard, Proceedings of Electronic Properties
  of Two-Dimensional Systems XI [Surf. Sci. (to be published)].

\bibitem{Pals} P. Pals and A. MacKinnon, preprint (12/31/96,
  cond-mat/9601155); preprint (12/31/96, cond-mat/9601156).

\bibitem{Amit} D. J. Amit, {\it Field Theory, the Renormalization
  Group, and Critical Phenomena}, 2nd ed. (World Scientific,
  Singapore, 1984).

\bibitem{Solyom} J.~Solyom, Adv. Phys. {\bf 28}, 201 (1979).

\bibitem{Negele} J. W. Negele and H. Orland, {\it Quantum
  Many-Particle Systems} (Addison-Wesley, Redwood City, 1988).

\bibitem{Grad} I. S. Gradshteyn and I. M. Ryzhik,
{\it Table of Integrals, Series, and Products}, 5th ed.,
edited by Alan Jeffrey (Academic, Boston, 1994), 933ff.



\end{references}
\end{document}